\documentclass[%
preprint,
amsmath,amssymb,
aps,
pra,
]{revtex4-2}

\usepackage{graphicx}
\usepackage{dcolumn}
\usepackage{bm}

\begin{document}


\title{Photonic Quantum Chromo-Dynamics}


\author{Shinichi Saito}
 \email{shinichi.saito.qt@hitachi.com}
\affiliation{Center for Exploratory Research Laboratory, Research \& Development Group, Hitachi, Ltd. Tokyo 185-8601, Japan.}

\date{\today}

\begin{abstract}
Vortexed photons with finite orbital angular momentum have a distinct mode profile with topological charge at the centre of the mode, while propagating in a certain direction.
Each mode with different topological charge of $m$ is orthogonal, in the sense that the overlap integral vanishes among modes with different values of $m$. 
Here, we theoretically consider a superposition state among 3 different modes with left- and right-vorticies and a Gaussian mode without a vortex.
These 3 states are considered to be assigned to different quantum states, thus, we have employed the $\mathfrak{su}(3)$ Lie algebra and the associated SU(3) Lie group to classify the photonic states.
We have calculated expectation values of 8 generators of the $\mathfrak{su}(3)$ Lie algebra, which should be observable, since the generators are Hermite.
We proposed to call these parameters as Gell-Mann parameters, named after the theoretical physicist, Murray Gell-Mann, who established Quantum Chromo-Dynamics (QCD) for quarks.
The Gell-Mann parameters are represented on the 8-dimensional hypersphere with its radius fixed due to the conservation law of the Casimir operator.
We have discussed a possibility to explore photonic QCD in experiments and classified SU(3) states to embed the parameters in SO(6) and SO(5).
\end{abstract}

\maketitle


\section{Introduction}

Lie algebra and Lie group \cite{Stubhaug02,Fulton04,Hall03,Pfeifer03,Georgi99,Cisowski22} were developed mathematically much earlier than the discoveries of quantum mechanics \cite{Dirac30,Baym69,Sakurai14,Georgi99,Weinberg05}.
The theory formulates general principles on how to classify various matrices with complex numbers ($\mathbb{C}$) and gives deep insights into the topological structure, underlying matrix calculations.
It covers quite wide areas and thus applicable to many fields in quantum physics, including various 2-level systems, described by the Spatial Unitary group of 2-dimension, SU(2) to understand polarisation, for example \cite{Stokes51,Poincare92,Jones41,Fano54,Baym69,Sakurai14,Max99,Jackson99,Yariv97,Gil16,Goldstein11,
Hecht17,Pedrotti07,Saito20a,Saito20b,Saito20c,Saito20d}.
Historically, however, the powerful mathematical features are not widely recognised in SU(2), since it is not so much complicated to deal with, even if we do not employ the knowledge of Lie algebra.
The situation has completely changed, once Murray Gell-Mann identified the underlying symmetries for composite elementary particles of baryons and mesons, establishing the Quantum Chromo-Dynamics (QCD) \cite{Gell-Mann61,Gell-Mann64,Ne'eman61,Pfeifer03,Sakurai67,Georgi99,Weinberg05}.
Lie algebra is now an indispensable tool for physics on elementary particles.
Here, we propose to introduce the framework of Lie algebra and Lie group to photonics, especially for exploring photonic analogue of QCD by utilising photonic orbital angular momentum.

\section{Summary of the $\mathfrak{su}(3)$ Lie algebra}

First, we summarise theoretically minimum knowledge on fundamental properties of the $\mathfrak{su}(3)$ Lie algebra, in order to make our discussions self-contained and clarify our notations.
We wish this will help photonic researchers, who are not familiar with the $\mathfrak{su}(3)$ algebra, to understand the idea to treat 3 orthogonal quantum states in an equal fashion.
It is far from a comprehensive summary, such that interested readers are encouraged to refer to excellent textbooks \cite{Stubhaug02,Fulton04,Hall03,Pfeifer03,Dirac30,Georgi99} .
Those who are familiar with the $\mathfrak{su}(3)$ algebra should skip this section.

The Lie algebra and the Lie group were mathematically developed as early as 1870s, without specific applications in physics \cite{Stubhaug02,Fulton04,Hall03,Pfeifer03,Cisowski22} . 
The first serious applications in physics found in elementary physics, leading to the discoveries of quarks \cite{Georgi99}.
Obviously, the $\mathfrak{su}(3)$ Lie algebra and more general representation theories are robust and they will be applicable to various cases. 
On the other hand, here, we are focussing on applications in photonics, and we will use this example to review fundamental characteristics of $\mathfrak{su}(3)$ algebra. 
Thus, we will lose the generality in our construction of the logic, but it will be easier to understand the concept and applications in higher dimensions will be straightforward.  

\subsection{Generators of the $\mathfrak{su}(3)$ algebra}
We consider 3 orthogonal quantum states, such as left- and right-vortexed states and the no-vortex state, which are described in the Hilbert space with 3 complex numbers, $\mathbb{C}^{3}$. 
We allow arbitrary mixing of these 3 states, realised by the superposition principle, and the wavefunction could be consider to be normalised to 1 or to the fixed number of photons, $N$, such that the radius of the complex sphere is fixed.
Consequently, the number of freedom is $2 \times 3 -1 =5$, and the Hilbert space is equivalent to the sphere of 5 dimensions, $S^5$.
In order to describe arbitrary rotational operations of the wavefunction in the Hilbert space, we need complex matrices of $3 \times 3$ for the SU(3) Lie group, which is realised by the exponential mapping form the $\mathfrak{su}(3)$ Lie algebra. 
The SU(3) forms a group, whose determinant must be unity, which corresponds to the traceless condition for the Lie algebra. 
Therefore, we need $3 \times 3 -1 =8$ bases, defined by 
\begin{eqnarray}
\hat{\lambda}_1
&=&
\begin{pmatrix}
0 & 1 & 0 \\
1 & 0 & 0 \\
0 & 0 & 0 
\end{pmatrix}
\nonumber
\\
\hat{\lambda}_2
&=&
\begin{pmatrix}
0 & -i & 0 \\
i & 0 & 0 \\
0 & 0 & 0 
\end{pmatrix}
\nonumber
\\
\hat{\lambda}_3
&=&
\begin{pmatrix}
1 & 0 & 0 \\
0 & -1 & 0 \\
0 & 0 & 0 
\end{pmatrix}
\nonumber
\\
\hat{\lambda}_4
&=&
\begin{pmatrix}
0 & 0 & 1 \\
0 & 0 & 0 \\
1 & 0 & 0 
\end{pmatrix}
\nonumber
\\
\hat{\lambda}_5
&=&
\begin{pmatrix}
0 & 0 & -i \\
0 & 0 & 0 \\
i & 0 & 0 
\end{pmatrix}
\nonumber
\\
\hat{\lambda}_6
&=&
\begin{pmatrix}
0 & 0 & 0 \\
0 & 0 & 1 \\
0 & 1 & 0 
\end{pmatrix}
\nonumber
\\
\hat{\lambda}_7
&=&
\begin{pmatrix}
0 & 0 & 0 \\
0 & 0 & -i \\
0 & i & 0 
\end{pmatrix}
\nonumber
\\
\hat{\lambda}_8
&=&
\frac{1}{\sqrt{3}}
\begin{pmatrix}
1 & 0 & 0 \\
0 & 1 & 0 \\
0 & 0 & -2 
\end{pmatrix}, 
\nonumber
\end{eqnarray}
which are all Hermite, $\hat{\lambda}_i^{\dagger}=\hat{\lambda}_i$ ($i=1,\cdots, 8$), implying their expectation values must be real and observable.
We can also uses the bases, $\hat{e}_i=\hat{\lambda}_i/2$, reflecting underlying the $\mathfrak{su}(2)$ symmetry among 2 orthogonal states. 
The bases satisfy the normalisation relationship for the trace, 
$
{\rm Tr}
\left(
\hat{\lambda}_i
\cdot
\hat{\lambda}_j
\right)
=2 \delta_{ij},
$ 
where $\delta_{ij}$ is the Kronecker delta.

\subsection{Commutation relationship}
The commutation relationship is obtained by straightforward calculation of basis matrices, and we obtain
\begin{eqnarray}
\left[
\hat{\lambda}_i
,
\hat{\lambda}_j
\right]
=
2 i
\sum_{k} 
C_{ijk}
\hat{\lambda}_k
,
\end{eqnarray}
where the structure constants, $C_{ijk}$, are listed in Table I.
$C_{ijk}$ is an asymmetric tensor, such that odd permutation of indices change its sign.
The most of commutation relationship involve only 1 term in the summation on the right hand side of the equation, similar to the $\mathfrak{su}(2)$ commutation relationship for spin.
On the other hand, we must account for 2 terms involved in equations 
\begin{eqnarray}
\left[
\hat{\lambda}_4
,
\hat{\lambda}_5
\right]
=
2 i
\left( 
\frac{1}{2}
\hat{\lambda}_3
+
\frac{\sqrt{3}}{2}
\hat{\lambda}_8
\right),
\end{eqnarray}
\begin{eqnarray}
\left[
\hat{\lambda}_6
,
\hat{\lambda}_7
\right]
=
2 i
\left( 
-
\frac{1}{2}
\hat{\lambda}_3
+
\frac{\sqrt{3}}{2}
\hat{\lambda}_8
\right).
\end{eqnarray}

We also confirm that we have 2 mutually commutable operators,
\begin{eqnarray}
\left[
\hat{\lambda}_1
,
\hat{\lambda}_8
\right]
=
\left[
\hat{\lambda}_2
,
\hat{\lambda}_8
\right]
=
\left[
\hat{\lambda}_3
,
\hat{\lambda}_8
\right]
=
0
\end{eqnarray}
while we see
\begin{eqnarray}
\left[
\hat{\lambda}_1
,
\hat{\lambda}_2
\right]
\neq 0
, \ 
\left[
\hat{\lambda}_1
,
\hat{\lambda}_3
\right]
\neq 0
, \ 
\left[
\hat{\lambda}_2
,
\hat{\lambda}_3
\right]
\neq 0.
\end{eqnarray}
Therefore, the rank 2 character of the $\mathfrak{su}(3)$ algebra is confirmed.
This is also evident that $\hat{\lambda}_3$ and $\hat{\lambda}_8$ are already diagonalised in our representation for the basis.

\begin{table}
\caption{\label{Table-I}
Structure constant of the commutation relationship, $[\hat{\lambda}_i,\hat{\lambda}_j ] = 2 i \sum_{k} C_{ijk} \hat{\lambda}_k$, in the $\mathfrak{su}(3)$ Lie algebra.
}
\begin{ruledtabular}
\begin{tabular}{cccccc}
& $i$ & $j$ & $k$ & $C_{ijk}$  & \\
\colrule
& 1 & 2 & 3 & 1& \\
& 1 & 4 & 7 & 1/2&  \\
& 1 & 5 & 6 & -1/2&  \\
& 2 & 4 & 6 & 1/2&  \\
& 2 & 5 & 7 & 1/2&  \\
& 3 & 4 & 5 & 1/2&  \\
& 3 & 6 & 7 & -1/2&  \\
& 4 & 5 & 8 & $\sqrt{3}/2$ & \\
& 6 & 7 & 8 & $\sqrt{3}/2$& 
\end{tabular}
\end{ruledtabular}
\end{table}

\subsection{Basis operators for the $\mathfrak{su}(2)$ algebra within the $\mathfrak{su}(3)$ algebra}

We are considering 3 orthogonal states for the $\mathfrak{su}(3)$ algebra, and we can pick up 2 states among 3 available states.
There are 3 ways to choose 2 states and each of the pairs of states will form the $\mathfrak{su}(2)$ Lie algebra.

For example, if we choose the first and the second states, corresponding to the left and the right vortexed states, we use bases
\begin{eqnarray}
\hat{e}_1^{(t)}
&=&
\frac{1}{2}
\hat{\lambda}_1
\\
\hat{e}_2^{(t)}
&=&
\frac{1}{2}
\hat{\lambda}_2
\\
\hat{e}_3^{(t)}
&=&
\frac{1}{2}
\hat{\lambda}_3,
\end{eqnarray}
for describing the $\mathfrak{su}(2)$ states, since they are equivalent to Pauli matricies, 
\begin{eqnarray}
\hat{\sigma}_1
&=&
\begin{pmatrix}
0 & 1  \\
1 & 0 
\end{pmatrix}
\\
\hat{\sigma}_2
&=&
\begin{pmatrix}
0 & -i  \\
i & 0 
\end{pmatrix}
\\
\hat{\sigma}_3
&=&
\begin{pmatrix}
1 & 0  \\
0 & -1 
\end{pmatrix}
,
\end{eqnarray}
if we neglect the third quantum state for the no-vortex state.
These operators, $\hat{e}_i^{(t)}$, were originally used for describing isospin for quarks \cite{Georgi99}.
For our applications, they will be useful to describe the rotation between the left and the right vortexed states.
The rotation corresponds to mixing the left and the right vortexed states, which will be described in the Poincar\'e sphere for vortices.

Another pair of states are made of the right vortexed state and the no-vortex state, whose bases are
\begin{eqnarray}
\hat{e}_1^{(u)}
&=&
\frac{1}{2}
\hat{\lambda}_6
\\
\hat{e}_2^{(u)}
&=&
\frac{1}{2}
\hat{\lambda}_7
\\
\hat{e}_3^{(u)}
&=&
\frac{1}{2}
\left(
-
\frac{1}{2}
\hat{\lambda}_3
+
\frac{\sqrt{3}}{2}
\hat{\lambda}_8
\right)
\\
&=&
\frac{1}{2}
\begin{pmatrix}
0 & 0 & 0 \\
0 & 1 & 0 \\
0 & 0 & -1 
\end{pmatrix}
.
\end{eqnarray}
Here, it is important to be aware that we can define a new vector operator of $\hat{e}_3^{(u)}$, for example, from $\hat{\lambda}_3$ and $\hat{\lambda}_8$, since they are basis vector operators in the $\mathfrak{su}(3)$ algebra, which forms a vector space. 
We cannot simply add components in the SU(3) Lie group, since SU(3) Lie group is not a vector space.
We see that $\hat{e}_3^{(u)}$ is normalised to be similar to $\hat{e}_3^{(t)}$, such that it is useful to consider SU(2) rotations by $\hat{e}_1^{(u)}$, $\hat{e}_2^{(u)}$, and $\hat{e}_3^{(u)}$.

Similarly, we consider the pair of states made of the left vortex state and the no-vortex state, whose bases are
\begin{eqnarray}
\hat{e}_1^{(v)}
&=&
\frac{1}{2}
\hat{\lambda}_4
\\
\hat{e}_2^{(v)}
&=&
\frac{1}{2}
\hat{\lambda}_5
\\
\hat{e}_3^{(v)}
&=&
\frac{1}{2}
\left(
\frac{1}{2}
\hat{\lambda}_3
+
\frac{\sqrt{3}}{2}
\hat{\lambda}_8
\right)
\\
&=&
\frac{1}{2}
\begin{pmatrix}
1 & 0 & 0 \\
0 & 0 & 0 \\
0 & 0 & -1 
\end{pmatrix}
\end{eqnarray}

These $\mathfrak{su}(2)$ commutation relationships are summarised as 
\begin{eqnarray}
\left[
\hat{e}_1^{(x)}
,
\hat{e}_2^{(x)}
\right]
=
i
\hat{e}_3^{(x)}
,
\end{eqnarray}
where $x=t,u, {\rm or} \  v$.
We have used small letters ($x=t, u, {\rm or} \  v$) for operators describing for a single quanta like a quark or a photon, and we will use capital letters for coherent states of photons  ($X=T, U, {\rm or} \  V$) under Bose-Einstein condensation, where macroscopic number, $N$, of photons are occupying the same state, latter of this paper.

The $\mathfrak{su}(3)$ algebra does not contain a non-trivial invariant group.
For example, we see
\begin{eqnarray}
\left[
\hat{e}_1^{(t)}
,
\hat{e}_4
\right]
=
-
\frac{1}{4}
\left[
\hat{\lambda}_1
,
\hat{\lambda}_4
\right]
=
-
\frac{1}{4}
i
\hat{\lambda}_7
=
-
\frac{1}{2}
\hat{e}_7
,
\end{eqnarray}
such that the internal $\mathfrak{su}(2)$ groups are connected by commutation relationships and the $\mathfrak{su}(2)$ algebra is not closed.

\subsection{Ladder operators}
We will utilise the Cartan-Dynkin formulation \cite{Georgi99} for describing the $\mathfrak{su}(3)$ states.
In the formalism, we consider to fix the quantisation axis rather than isotropic to all directions in the $\mathfrak{su}(2)$ algebra, and consider ladder operators for rising and lowering the qnautm number along the quantisation axis \cite{Georgi99,Sakurai14}.
More specifically, we define 
\begin{eqnarray}
\hat{t}_{\pm}
&=&
\frac{1}{2}
\left(
\hat{\lambda}_1
\pm
i
\hat{\lambda}_2
\right)
\\
\hat{t}_{3}
&=&
\frac{1}{2}
\hat{\lambda}_3,
\end{eqnarray}
where $\hat{t}_{3}$ stands for the operator of the $z$ component of the rotationally-symmetric $\mathfrak{su}(2)$ operator $\hat{\bf t}=(\hat{t}_{1},\hat{t}_{2},\hat{t}_{3})$, and $\hat{t}_{+}$ and $\hat{t}_{-}$ are the rising and the lowering operators, respectively, to increment and decrement the quantum number for $\hat{t}_{3}$.
We can use $\hat{t}_{3}$ and $\hat{t}_{\pm}$ instead of $\hat{e}_i^{(t)}$ ($i=1,2,3$), and their commutation relationships become 
\begin{eqnarray}
\left[
\hat{t}_{3}
,
\hat{t}_{\pm}
\right]
&=&
\pm
\hat{t}_{\pm}
\\
\left[
\hat{t}_{+}
,
\hat{t}_{-}
\right]
&=&
2
\hat{t}_{3}
.
\end{eqnarray}
For applications to isospin, $\hat{t}_{3}$ gives the fixed isospin value ($t_3$) for each elementary particle, such as a proton ($t_3=1/2$) and a neutron ($t_3=-1/2$), a deuterium (D, $t_3=0$), and a tritium (T, $t_3=1/2$).
In elementary particle physics, a superposition state between a proton and a neutron, for example, is not realised due to the superselection rule \cite{Georgi99}, since the superposition state between different charged states is prohibitted.
On the other hand, for applications to vortices, we can safely consider the superposition state between the left and the right vortexed states \cite{Padgett99,Saito21f}, such that we can consider arbitrary mixing of left and right vortexed states with arbitrary phase between them.
The ladder operators $\hat{t}_{\pm}$ correspond to changing the topological charge at the centre of the vortices for changing its orbital angular momentum from the left to the right circulation or {\it vice versa}.

Similarly, we consider the rising and the lowering ladder operators, $\hat{u}_{+}$ and $\hat{u}_{-}$, respectively, for the superposition state between the right vortexed and the no-vortex states, and the $z$ component of the rotationally-symmetric $\mathfrak{su}(2)$ operator $\hat{\bf u}=(\hat{u}_{1},\hat{u}_{2},\hat{u}_{3})$
\begin{eqnarray}
\hat{u}_{\pm}
&=&
\frac{1}{2}
\left(
\hat{\lambda}_6
\pm
i
\hat{\lambda}_7
\right)
\\
\hat{u}_{3}
&=&
\frac{1}{4}
\left(
-
\hat{\lambda}_3
+
\sqrt{3}
\hat{\lambda}_8
\right),
\end{eqnarray}
whose commutation relationships become
\begin{eqnarray}
\left[
\hat{u}_{3}
,
\hat{u}_{\pm}
\right]
&=&
\pm
\hat{u}_{\pm}
\\
\left[
\hat{u}_{+}
,
\hat{u}_{-}
\right]
&=&
2
\hat{u}_{3}
.
\end{eqnarray}

For the mixing of the left-vortexed and no-vortex states, the corresponding ladder operators $\hat{v}_{+}$ and $\hat{v}_{-}$, and the $z$ component of the rotationally-symmetric $\mathfrak{su}(2)$ operator $\hat{\bf v}=(\hat{v}_{1},\hat{v}_{2},\hat{v}_{3})$ become
\begin{eqnarray}
\hat{v}_{\pm}
&=&
\frac{1}{2}
\left(
\hat{\lambda}_4
\pm
i
\hat{\lambda}_5
\right)
\\
\hat{v}_{3}
&=&
\frac{1}{4}
\left(
\hat{\lambda}_3
+
\sqrt{3}
\hat{\lambda}_8
\right)
,
\end{eqnarray}
whose commutation relationships become
\begin{eqnarray}
\left[
\hat{v}_{3}
,
\hat{v}_{\pm}
\right]
&=&
\pm
\hat{v}_{\pm}
\\
\left[
\hat{v}_{+}
,
\hat{v}_{-}
\right]
&=&
2
\hat{v}_{3}
.
\end{eqnarray}

Here, we have defined 9 operators for ladders and the quantisation components of the 3 sets of $\mathfrak{su}(2)$ operators $(\hat{\bf t},\hat{\bf u},\hat{\bf v})$, while only 8 bases are required for the $\mathfrak{su}(3)$ algebra, due to the traceless requirement.
Consequently, we have obtained 1 identity
\begin{eqnarray}
\hat{v}_{3}
=
\hat{u}_{3}
+
\hat{t}_{3}
,
\end{eqnarray}
which must be met for all states.
This means only 2 quantum numbers are independently chosen, regardless of apparent 3 sets of $\mathfrak{su}(2)$ states, which is in fact consistent with the rank-2 nature of the $\mathfrak{su}(3)$ algebra.

\subsection{Hypercharge and topological charge}
As we have reviewed above, we can pick up 2 quantum operators from 3 operators, $\hat{t}_3$, $\hat{u}_3$, and $\hat{v}_3$, for describing the $\mathfrak{su}(3)$ quantum states.
If we choose $\hat{t}_3$, we can choose $\hat{u}_3$ or $\hat{v}_3$.
Alternatively, we can consider the superposition state, made of both $\hat{u}_3$ and$\hat{v}_3$, whose $z$ component becomes
\begin{eqnarray}
\hat{\lambda}_8
=
\frac{2}{\sqrt{3}}
\left(
\hat{u}_3
+
\hat{v}_3
\right)
.
\end{eqnarray}

Equivalently, we can define the hypercharge operator \cite{Georgi99}, as
\begin{eqnarray}
\hat{y}
&=&
\frac{1}{\sqrt{3}}
\hat{\lambda}_8
\nonumber
\\
&=&
\frac{2}{3}
\left(
\hat{u}_3
+
\hat{v}_3
\right)
=
\frac{4}{3}
\hat{u}_3
+
\frac{2}{3}
\hat{t}_3
=
\frac{4}{3}
\hat{v}_3
-
\frac{2}{3}
\hat{t}_3
,
\end{eqnarray}
which was indispensable to understand quarks, their 2-body (3-body) compounds of meson, and the 3-body compounds of baryons.
It commutes with the other 3 operators, 
\begin{eqnarray}
\left[
\hat{y}
,
\hat{t}_{3}
\right]
=
\left[
\hat{y}
,
\hat{u}_{3}
\right]
=
\left[
\hat{y}
,
\hat{v}_{3}
\right]
=
0
,
\nonumber
\end{eqnarray}
meaning that hypercharge could be the simultaneous quantum number with the other parameter.
Therefore, we expect that the eigenstate is labelled by the quantum numbers, $t_{3}$, $u_3$, $v_3$ with $y=(u_3+v_3)/3$, to satisfy
\begin{eqnarray}
\hat{y}
\left |
t_{3}, u_3, v_3
\right \rangle
=
y
\left |
t_{3}, u_3, v_3
\right \rangle
=
y
\left |
t_{3}, y
\right \rangle
\nonumber
\end{eqnarray}
We also confirm the identity, 
\begin{eqnarray}
\left[
\hat{y}
,
\hat{t}_{\pm}
\right]
&=&
\frac{2}{3}
\left[
\hat{u}_3
,
\hat{t}_{\pm}
\right]
+
\frac{2}{3}
\left[
\hat{v}_3
,
\hat{t}_{\pm}
\right]
=0
,
\nonumber
\end{eqnarray}
which ensures 
\begin{eqnarray}
\hat{y}
\left(
\hat{t}_{\pm}
\left |
t_{3}, y
\right \rangle
\right)
&=&
y
\left(
\hat{t}_{\pm}
\left |
t_{3}, y
\right \rangle
\right)
,
\nonumber
\end{eqnarray}
meaning that the application of ladder operations by $\hat{t}_{\pm}$ will preserve the hypercharge, $y$.

On the other hand, we find
\begin{eqnarray}
\left[
\hat{y}
,
\hat{u}_{\pm}
\right]
=
\frac{2}{3}
\left[
\hat{u}_3
+
\hat{v}_3
,
\hat{u}_{\pm}
\right]
=
\pm
\hat{u}_{\pm}
,
\nonumber
\end{eqnarray}
which leads
\begin{eqnarray}
\hat{y}
\left(
\hat{u}_{\pm}
\left |
t_{3}, y
\right \rangle
\right)
=
(y \pm 1)
\left(
\hat{u}_{\pm}
\left |
t_{3}, y
\right \rangle
\right)
,
\end{eqnarray}
which means $\hat{u}_{+}$ increments $y$ and $\hat{u}_{-}$ decrements $y$, respectively. 
We also confirm the same rule for $\hat{v}_{\pm}$ as
\begin{eqnarray}
\left[
\hat{y}
,
\hat{v}_{\pm}
\right]
=
\frac{2}{3}
\left[
\hat{u}_3
+
\hat{v}_3
,
\hat{v}_{\pm}
\right]
=
\pm
\hat{v}_{\pm}
,
\nonumber
\end{eqnarray}
which corresponds to
\begin{eqnarray}
\hat{y}
\left(
\hat{v}_{\pm}
\left |
t_{3}, y
\right \rangle
\right)
=
(y \pm 1)
\left(
\hat{v}_{\pm}
\left |
t_{3}, y
\right \rangle
\right)
.
\end{eqnarray}

Finally, we obtain the fundamental multiplets (Fig. 1), given by 3 states
\begin{eqnarray}
| \psi_1 \rangle
&=&
\left | 
t_3= \frac{1}{2}, 
t_8= \frac{1}{3}\frac{\sqrt{3}}{2} 
\right \rangle
=
\begin{pmatrix}
1 \\
0 \\
0 
\end{pmatrix}
\nonumber
\\
| \psi_2 \rangle
&=&
\left | 
t_3= -\frac{1}{2}, 
t_8= \frac{1}{3}\frac{\sqrt{3}}{2} 
\right \rangle
=
\begin{pmatrix}
0 \\
1 \\
0 
\end{pmatrix}
\nonumber
\\
| \psi_3 \rangle
&=&
\left | 
t_3= 0, 
t_8= -\frac{2}{3}\frac{\sqrt{3}}{2} 
\right \rangle
=
\begin{pmatrix}
0 \\
0 \\
1 
\end{pmatrix}
.
\nonumber
\end{eqnarray}
An arbitrary quantum state can be generated by mixing these 3 states upon the superposition principle: multiplying complex numbers to fundamental ket states and summing up.
In a standard matrix formulation of quantum mechanics, a general state is given by a row of 3 complex numbers.

For quarks, there is an identity relationship between hypercharge and charge, q, as
\begin{eqnarray}
q=t_3 + \frac{1}{2}y
.
\end{eqnarray}
Therefore, one can use charge instead of hypercharge for an alternative quantum number.

For our applications to photonic orbital angular momentum, we consider superposition states between left- and right vortexed states and no-vortex state. 
We use the orbital angular momentum along the quantisation axis, $z$, which is the direction of the propagation, as the first quantum number, instead of the isospin of $t_3$.
For the second quantum number, instead of hypercharge, we choose the topological charge, defined by
\begin{eqnarray}
q_{\rm t}=y+\frac{2}{3}
,
\end{eqnarray}
which becomes 0 for the no-vortex state and 1 for both left- and right-vortexed states.
The topological charge corresponds to the winding number of the mode at the core, propagating along a certain $z$ direction.
It is also linked to the magnitude of photonic orbital angular momentum.
In this paper, we only consider vorticies with the winding number of 1 or 0, but it will be straightforward to extend our discussions to higher order states.

\begin{figure}[h]
\begin{center}
\includegraphics[width=8cm]{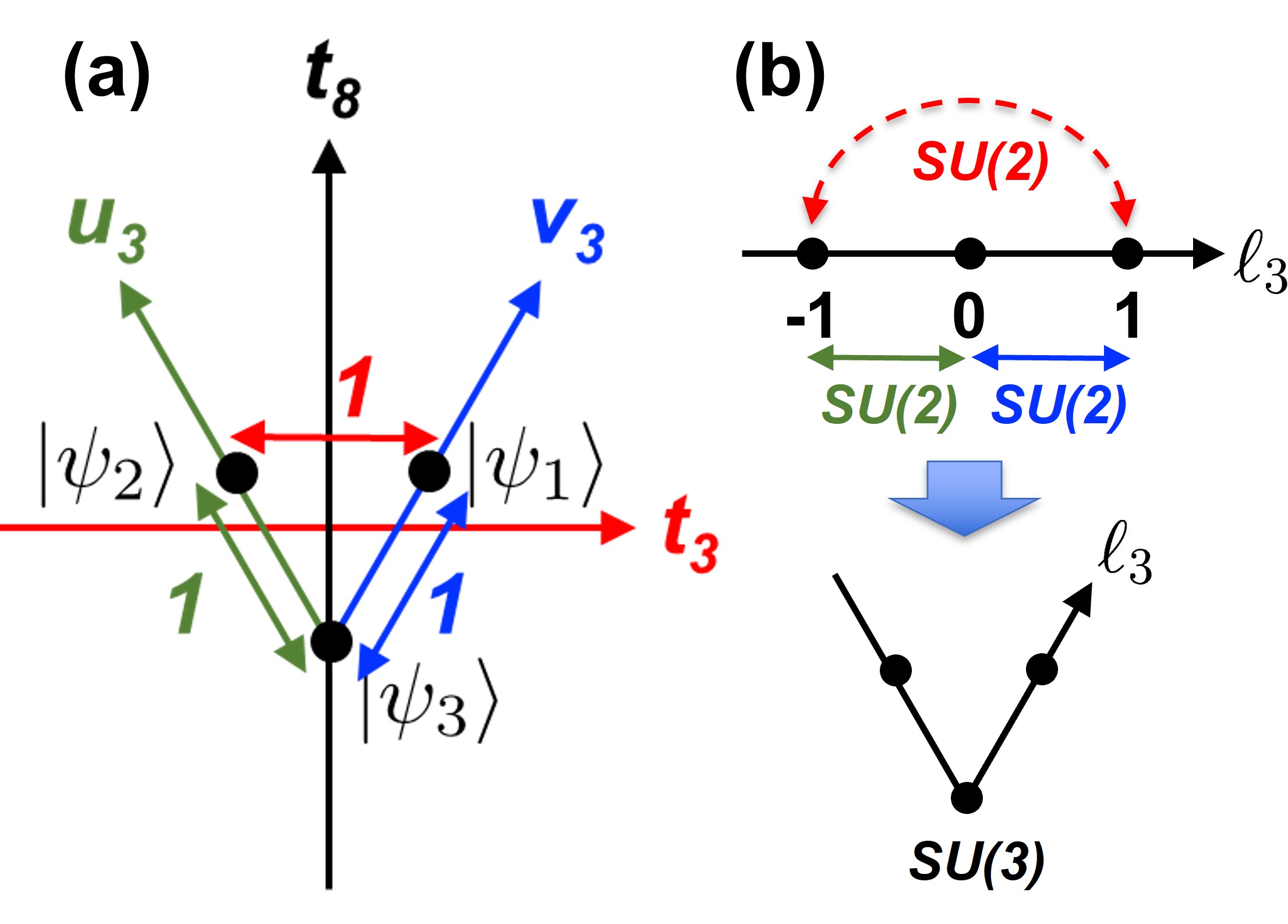}
\caption{
SU(3) states with photonic orbital angular momentum.
(a) Fundamental multiplet of the $\mathfrak{su}(3)$ algebra.
Fundamental basis states of $|\psi_1 \rangle$, $|\psi_2 \rangle$, and $|\psi_3 \rangle$ are shown on the $(t_3, t_8)$ plane, characterised by their quantum numbers.
$t_3$ is known as isospin for quarks.
The states are shown by points, given by the eigenvalues, which are separated by the same distance and form an equilateral triangle as their topology, implying the states are treated in an equal footing in the $\mathfrak{su}(3)$ algebra. 
We can use $u_3$, $v_3$, or hyperchrage of $y=2(u_3+v_3)/3$ instead of $t_8$, but only 2 vectors are required to span the $t_3$-$t_8$ plane due to the rank 2 character of the $\mathfrak{su}(3)$ algebra.
(b) Bending of the quantisation axis of SU(2) to form SU(3) states.
The quantum number ($\ell_3$) of orbital angular momentum along the quantisation axis is usually characterised by SU(2) states, as shown by the upper diagram.
By allowing the SU(2) rotation between left- and right-vortexed states, we will effectively bend the $\ell_3$ to realise the superposition state. 
By combining superposition states with no-vortex state, we will mix the 3 orthogonal states to realise SU(3) states.
}
\end{center}
\end{figure}

\subsection{Casimir Operators}
There are other conservative properties in the $\mathfrak{su}(3)$ algebra.
We define a Casimir operator, 
\begin{eqnarray}
\hat{C}_1
&=&
\frac{1}{4}
\sum_{i=1}^{n}
\hat{\lambda}_i^2
=
\sum_{i=1}^{n}
\hat{e}_i^2
.
\end{eqnarray}

We calculate the commutation relationship, 
\begin{eqnarray}
\left [ 
\hat{C}_1
,
\hat{\lambda}_j
\right ]
&=&
\frac{1}{4}
\sum_{i=1}^{n}
\left [ 
\hat{\lambda}_i^2
,
\hat{\lambda}_j
\right ]
\nonumber
\\
&=&
\frac{1}{4}
\sum_{i=1}^{n}
\left ( 
\hat{\lambda}_i^2
\hat{\lambda}_j
-
\hat{\lambda}_j
\hat{\lambda}_i^2
\right )
\nonumber
\\
&=&
\frac{1}{4}
\sum_{i=1}^{n}
\left ( 
\hat{\lambda}_i(
\hat{\lambda}_i
\hat{\lambda}_j)
-
(\hat{\lambda}_j
-
\hat{\lambda}_i)
\hat{\lambda}_i
\right )
\nonumber
\\
&=&
\frac{1}{4}
\sum_{i=1}^{n}
\left ( 
\hat{\lambda}_i
[\hat{\lambda}_i,\hat{\lambda}_j]
-
[\hat{\lambda}_j,\hat{\lambda}_i]
\hat{\lambda}_i
\right )
\nonumber
\\
&=&
\frac{2i}{4}
\sum_{ik}
(
C_{ijk}
\hat{\lambda}_i
\hat{\lambda}_k
-
C_{jik}
\hat{\lambda}_k
\hat{\lambda}_i
)
\nonumber
\\
&=0,
\end{eqnarray}
where we have changed the dummy indices at the last line as 
$\sum_{ik}C_{jik}
\hat{\lambda}_k
\hat{\lambda}_i
=
\sum_{ki}
C_{jki}
\hat{\lambda}_i
\hat{\lambda}_k
=
\sum_{ik}
C_{ijk}
\hat{\lambda}_i
\hat{\lambda}_k$
.
Therefore, the Casimir operator, $\hat{C}_1$, obtains the simultaneous eigenstate with the rank-2 states for $\hat{\lambda}_i$.
In fact, we see, from direct calculations, 
\begin{eqnarray}
\hat{C}_1=
\begin{pmatrix}
1+\frac{1}{3} & 0 & 0 \\
0 & 1+\frac{1}{3} & 0 \\
0 & 0 & 1+\frac{1}{3} 
\end{pmatrix}
=
\frac{4}{3},
\end{eqnarray}
which means that $\hat{C}_1$ is actually constant for $\mathfrak{su}(3)$ states. 
Here, it is obvious that we have abbreviated the unit matrix of $3\times 3$, ${\bf 1}_3$, multiplied with 
$\frac{4}{3}=\frac{4}{3}{\bf 1}_3$, for simplicity.

There is another Casimir operator, defined by 
\begin{eqnarray}
\hat{C}_2
&=&
\sum_{ijk}
D_{ijk}
\hat{t}_i
\hat{t}_j
\hat{t}_k
\nonumber
\\
&=&
\frac{1}{8}
\sum_{ijk}
D_{ijk}
\hat{\lambda}_i
\hat{\lambda}_j
\hat{\lambda}_k
,
\end{eqnarray}
where $D_{ijk}$ is a symmetric tensor, as defined in the anti-commutation relationship, below.
We see that $\hat{C}_2$ is also constant in the $\mathfrak{su}(3)$ algebra, such that the commutation relationship, 
\begin{eqnarray}
\left [
\hat{C}_2,
\hat{\lambda}_i
\right ]
=0,
\end{eqnarray}
vanishes.

\subsection{Anti-Commutation relation}
We also obtain the anti-commutation relationship as 
\begin{eqnarray}
\left \{
\hat{\lambda}_i
,
\hat{\lambda}_j
\right \}
&=&
\frac{4}{3}\delta_{ij}
+
2
\sum_{k=1}^{8}
D_{ijk}
\hat{\lambda}_k
,
\end{eqnarray}
which is equivalent to
\begin{eqnarray}
\left \{
\hat{e}_i
,
\hat{e}_j
\right \}
&=&
\frac{1}{3}\delta_{ij}
+
\sum_{k=1}^{8}
D_{ijk}
\hat{e}_k
,
\end{eqnarray}
where we have abbreviated ${\bf 1}_3$ in $\frac{4}{3}\delta_{ij}=\frac{4}{3}\delta_{ij}{\bf 1}_3$, as before.
The symmetric tensor, $D_{ijk}$, is shown in Table II.

By multiply $\hat{\lambda}_k$ to the anti-commutation relationship, we obtain
\begin{eqnarray}
\left \{
\hat{\lambda}_i
,
\hat{\lambda}_j
\right \}
\hat{\lambda}_k
&=&
\frac{4}{3}\delta_{ij}
\hat{\lambda}_k
+
2
D_{ijk^{\prime}}
\hat{\lambda}_{k^{\prime}}
\hat{\lambda}_k
.
\nonumber
\end{eqnarray}
We take the trace of the equation, while using ${\rm Tr}(\hat{\lambda}_i)=0$ and ${\rm Tr}(\hat{\lambda}_i \hat{\lambda}_j)=2 \delta_{ij}$, we obtain
\begin{eqnarray}
D_{ijk}
=
\frac{1}{4}
{\rm Tr}
\left (
\left \{
\hat{\lambda}_i
,
\hat{\lambda}_j
\right \}
\hat{\lambda}_k
\right)
.
\end{eqnarray}
Similarly, we also obtain
\begin{eqnarray}
C_{ijk}
=
\frac{1}{4i}
{\rm Tr}
\left(
\left[
\hat{\lambda}_i
,
\hat{\lambda}_j
\right]
\hat{\lambda}_k
\right)
\end{eqnarray}
from the commutation relationship.

\begin{table}
\caption{\label{Table-II}
Structure constant of the anti-commutation relationship, $\{\hat{\lambda}_i,\hat{\lambda}_j \} = 4\delta_{ij}/3+2  \sum_{k} D_{ijk} \hat{\lambda}_k$, in the $\mathfrak{su}(3)$ Lie algebra.}
\begin{ruledtabular}
\begin{tabular}{cccccc}
&$i$ & $j$ & $k$ & $D_{ijk}$ & \\
\colrule
&1 & 1 & 8 & $1/\sqrt{3}$ & \\
&1 & 4 & 6 & 1/2  & \\
&1 & 5 & 7 & 1/2  & \\
&2 & 2 & 8 & $1/\sqrt{3}$  & \\
&2 & 4 & 7 & -1/2  & \\
&3 & 5 & 6 & 1/2  & \\
&3 & 3 & 8 & $1/\sqrt{3}$  & \\
&3 & 4 & 4 & 1/2  & \\
&3 & 5 & 5 & 1/2 & \\
&3 & 6 & 6 & -1/2  & \\
&3 & 7 & 7 & -1/2  & \\
&4 & 4 & 8 & -$1/(2\sqrt{3})$  & \\
&5 & 5 & 8 & -$1/(2\sqrt{3})$  & \\
&6 & 6 & 8 & -$1/(2\sqrt{3})$  & \\
&7 & 7 & 8 & -$1/(2\sqrt{3})$  & \\
&8 & 8 & 8 & -$1/\sqrt{3}$  & 
\end{tabular}
\end{ruledtabular}
\end{table}

Finally, we will show
\begin{eqnarray}
\hat{C}_2
&=
\hat{C}_1
\left(
2\hat{C}_1
-\frac{11}{6}
\right)
=
\frac{10}{9}
.
\nonumber
\end{eqnarray}
To prove the identity, we use 
\begin{eqnarray}
\left \{
\hat{\lambda}_i
,
\hat{\lambda}_j
\right \}
&=
\hat{\lambda}_i
\hat{\lambda}_j
+
\hat{\lambda}_j
\hat{\lambda}_i
\nonumber
\end{eqnarray}
and
\begin{eqnarray}
\left [
\hat{\lambda}_i
,
\hat{\lambda}_j
\right ]
&=
\hat{\lambda}_i
\hat{\lambda}_j
-
\hat{\lambda}_j
\hat{\lambda}_i
.
\nonumber
\end{eqnarray}
By adding these equations, we obtain 
\begin{eqnarray}
\left \{
\hat{\lambda}_i
,
\hat{\lambda}_j
\right \}
+
\left [
\hat{\lambda}_i
,
\hat{\lambda}_j
\right ]
&=
2
\hat{\lambda}_i
\hat{\lambda}_j
,
\nonumber
\end{eqnarray}
which becomes
\begin{eqnarray}
\frac{4}{3}\delta_{ij}
+
2
D_{ijk}
\hat{\lambda}_k
+
2i
C_{ijk}
\hat{\lambda}_k
&=
2
\hat{\lambda}_i
\hat{\lambda}_j
\nonumber
\end{eqnarray}
from commutation and anti-commutation relationships.
Then, we multiply a factor of $\hat{\lambda}_i \hat{\lambda}_j$ and sum up to obtain
\begin{eqnarray}
\sum_{ij}
\hat{\lambda}_i^2
\hat{\lambda}_j^2
&=&
\frac{2}{3}
\sum_{i}
\hat{\lambda}_i^2
+
\sum_{ijk}
D_{ijk}
\hat{\lambda}_i
\hat{\lambda}_j
\hat{\lambda}_k
\nonumber
\\
&&-
i
\sum_{ijk}
C_{ijk}
\hat{\lambda}_i
\hat{\lambda}_j
\hat{\lambda}_k
,
\end{eqnarray}
where the last term becomes
\begin{eqnarray}
\sum_{ijk}
C_{ijk}
\hat{\lambda}_i
\hat{\lambda}_j
\hat{\lambda}_k
&=&
\frac{1}{2}
\sum_{ijk}
\left(
C_{ijk}
\hat{\lambda}_i
\hat{\lambda}_j
\hat{\lambda}_k
+
C_{ikj}
\hat{\lambda}_i
\hat{\lambda}_k
\hat{\lambda}_j
\right)
\nonumber
\\
&=&
\frac{1}{2}
\sum_{ijk}
\left(
C_{ijk}
\hat{\lambda}_i
\hat{\lambda}_j
\hat{\lambda}_k
-
C_{ijk}
\hat{\lambda}_i
\hat{\lambda}_k
\hat{\lambda}_j
\right)
\nonumber
\\
&=&
\frac{1}{2}
\sum_{ijk}
C_{ijk}
\hat{\lambda}_i
\left(
\hat{\lambda}_j
\hat{\lambda}_k
-
\hat{\lambda}_k
\hat{\lambda}_j
\right)
\nonumber
\\
&=&
\frac{1}{2}
\sum_{ijk}
C_{ijk}
\hat{\lambda}_i
\left[
\hat{\lambda}_j
,
\hat{\lambda}_k
\right]
\nonumber
\\
&=&
i
\sum_{ijk}
C_{ijk}
C_{jkl}
\hat{\lambda}_i
\hat{\lambda}_l
\nonumber
\\
&=&
i
\sum_{ijk}
C_{jki}
C_{jkl}
\hat{\lambda}_i
\hat{\lambda}_l
\nonumber
\\
&=&
3i
\sum_{i}
\hat{\lambda}_i^2
,
\nonumber
\end{eqnarray}
which leads 
\begin{eqnarray}
\sum_{ij}
\hat{\lambda}_i^2
\hat{\lambda}_j^2
=
\sum_{ijk}
D_{ijk}
\hat{\lambda}_i
\hat{\lambda}_j
\hat{\lambda}_k
+
\left(
\frac{2}{3}+3
\right)
\sum_{i}
\hat{\lambda}_i^2
.
\nonumber
\end{eqnarray}
Therefore, we obtain
\begin{eqnarray}
\hat{C}_2
&=&
\hat{C}_1
\left(
2\hat{C}_1
-\frac{11}{6}
\right)
=
\frac{10}{9},
\end{eqnarray}
which is in fact constant under the $\mathfrak{su}(3)$ algebra.

\section{SU(3) state for vortexed modes}
\subsection{Gell-Mann Hypersphere}

Now, we are ready for discussing how to classify the superposition state between left- and right-vortexed states and no-vortex state using the $\mathfrak{su}(3)$ algebra and the SU(3) Lie group.
We assume Laguerre-Gauss modes with topological charge of $q_{\rm t}=1$ for both left- and right-vortexed states \cite{Allen92,Padgett99,Milione11,Naidoo16,Liu17,Erhard18,Saito21f,Andrews21,Angelsky21}, for simplicity.

Here, our main idea is to assign the 3 states of vortexed modes and no-vortex mode to orthogonal states of the SU(3) states (Fig. 1).
The most important part of the vortexed modes for orbital angular momentum is its azimuthal ($\phi$) dependence \cite{Allen92}, i.e., the wavefunction of the ray with orbital angular momentum of $m$ is given by
\begin{eqnarray}
\langle \phi|m \rangle = {\rm e}^{i m \phi}
,
\end{eqnarray}
which is orthogonal each other for states with different $m$ in a sense,
\begin{eqnarray}
\langle m^{\prime} | m \rangle = 
\int_0^{2 \pi} \frac{d \phi}{2 \pi}
{\rm e}^{i (m-m^{\prime} \phi}
=\delta_{m,m^{\prime}}
.
\end{eqnarray}
This means that the modes with different orbital angular momentum could be treated as orthogonal states as quantum mechanical states.
For our consideration and notation \cite{Saito20a,Saito20b,Saito20c,Saito20d,Saito20e,Saito21f,Saito22g,Saito22h,Saito22i}, we will assign left- and right vortexed states as $|{\rm  L} \rangle = |1 \rangle= |\psi_1 \rangle $ and $|{\rm  R} \rangle = |-1 \rangle= |\psi_2 \rangle$, respectively, and the no-vortex Gaussian state $|{\rm  O} \rangle = |0 \rangle= |\psi_3 \rangle$.

We also assume that all modes have the same polarisation state, such that our SU(3) state is polarised.
Later, we will consider the polarisation degree of freedom, which comes from the SU(2) spin of photons \cite{Stokes51,Poincare92,Jones41,Fano54,Baym69,Sakurai14,Max99,Jackson99,Yariv97,Gil16,Goldstein11,
Hecht17,Pedrotti07,Saito20a,Saito20b,Saito20c,Saito20d}, such that we explore the photonic states with the SU(2) $ \times$ SU(3) symmetry.

We consider coherent ray of photons emitted from a laser source \cite{Yariv97,Gil16,Goldstein11,Saito20a,Saito21f,Saito22g,Saito22h}, such that a macroscopic number of photons per second, $N$, is passing through the cross section of the ray.
We use capital letters to describe macroscopic observables and expectation values, such as photonic orbital angular momentum 
\cite{Allen92,Padgett99,Milione11,Naidoo16,Liu17,Erhard18,Saito21f,Andrews21,Angelsky21} 
\begin{eqnarray}
\hat{L}_{i}
=
\hbar
N
\hat{\ell}_i
=
\hbar
N
\hat{\lambda}_i
,
\end{eqnarray}
where $\hbar=h/(2\pi)$ is the Dirac constant, defined by the plank constant ($h$), divided by $2\pi$, 
while small letters are used for single quantum operator or a normalised parameter, such as a normalised orbital angular momentum operator, 
\begin{eqnarray}
\hat{\ell}_i
=
\hat{\lambda}_i
,
\end{eqnarray}
for $i=1$, $2$, and $3$.

There is a factor of 2 difference in definition between the orbital angular momentum operator $\hat{\ell}_i$ and the isospin operator of $\hat{t}_3$, but it would be more appropriate to use $\hat{\ell}_i$ for photonic vortices, since the orbital angular momentum is quantised in the unit of $\hbar$ \cite{Allen92,Padgett99,Saito20a,Saito20b,Saito20c,Saito21f}.

First, let us review the SU(2) coupling between left- and right- vortexed states \cite{Padgett99,Saito21f}.
We consider the following state,
\begin{eqnarray}
|
\theta_l,
\phi_l
\rangle
=
\left (
  \begin{array}{c}
    {\rm e}^{-i\frac{\phi_l}{2}}  \cos \left( \frac{\theta_l}{2} \right)    \\
    {\rm e}^{+i\frac{\phi_l}{2}}  \sin \left( \frac{\theta_l}{2} \right)  \\
    0
  \end{array}
\right)
,
\end{eqnarray}
where the amplitudes of left- and right-vortex states are controlled by the polar angle of $\theta_l$ and the phase is defined by $\phi_l$.
We can realise this state by an exponential map from $\mathfrak{su}(3)$ Lie algebra to the SU(3) Lie group
\begin{eqnarray}
\hat{\mathcal D}_2 \left( \theta_l \right)
&=&\exp \left( -\frac{i \hat{\lambda}_2 \theta_l}{2} \right)
\\
&=&
\begin{pmatrix}
\cos \left( \frac{\theta_l}{2} \right)  &  - \sin \left( \frac{\theta_l}{2} \right)  & 0 \\
\sin \left( \frac{\theta_l}{2} \right)  &     \cos \left( \frac{\theta_l}{2} \right)  & 0\\
0 & 0 & 0 
\end{pmatrix}
,
\end{eqnarray}
which is a phase-shifter with its fast axis rotated for $\pi/4$ from the horizontal axis \cite{Saito20a}, 
together with another exponential map of 
\begin{eqnarray}
\hat{\mathcal D}_3 \left( \phi_l \right)
&=&\exp \left( -\frac{i \hat{\lambda}_3 \phi_l}{2} \right)
\\
&=&
\begin{pmatrix}
\exp \left( -i\frac{\phi_l}{2} \right) & 0 & 0 \\
0 & \exp \left( i\frac{\phi_l}{2} \right)   & 0 \\
0 & 0 & 0 
\end{pmatrix}
,
\end{eqnarray}
which is a rotator.
We apply these operators to a unit vector, $|\psi_1\rangle$, to confirm the SU(2) state 
\begin{eqnarray}
|
\theta_l,
\phi_l
\rangle
&=&
\hat{\mathcal D}_3 \left( \phi_l \right)
\hat{\mathcal D}_2 \left( \theta_l \right)
|\psi_1\rangle
,
\end{eqnarray}
made of left- and right-vortexed states.

By calculating a standard quantum mechanical average from $|\theta_l, \phi_l \rangle$, 
\begin{eqnarray}
\ell_i
=
\lambda_i
=
\langle
\hat{\ell}_i
\rangle
=
\langle
\hat{\lambda}_i
\rangle
=
\langle \theta_l,\phi_l |
\hat{\lambda}_i
| \theta_l,\phi_l \rangle
,
\end{eqnarray}
for $i=1$, $2$, and $3$, respectively, we obtain
\begin{eqnarray}
\lambda_1
&=&  
\sin \left( \theta_l \right)    
\cos \left( \phi_l \right)    
\\
\lambda_2
&=&
\sin \left( \theta_l \right)    
\sin \left( \phi_l \right)  
\\
\lambda_3
&=&
\cos \left( \theta_l \right)    
.
\end{eqnarray}
Thus, the SU(2) states between left- and right-vortexed states could be shown on the Poincar\'e sphere for orbital angular momentum \cite{Padgett99,Milione11,Liu17,Saito21f}.
Usually, the rotational symmetry and corresponding orbital angular momentum are considered by the SU(2) symmetry, since the states between $|{\rm  L} \rangle = |1 \rangle$ and  $|{\rm  R} \rangle = |-1 \rangle$ cannot be transferred by the change of $\Delta m=\pm 1$, and instead, $\Delta m=\pm2$ is required.
This could be achieved by using a spiral phase plate \cite{Golub07} with topological charge of $m=2$.
Alternatively, it is possible to make a superposition state between $|{\rm  L} \rangle = |1 \rangle$ and  $|{\rm  R} \rangle = |-1 \rangle$, and SU(2) states could be realised by controlling the amplitudes and the phases \cite{Padgett99,Saito21f}.
For our considerations in the SU(3) states, this corresponds to bend the quantization axis, $\hat{\ell}_3$, for allowing 3 states to couple each other (Fig. 1(b)).

Now, we proceed to consider coupling between the no-vortex state and left- or right-vortex states.
This corresponds to change the hypercharge and the topological charge.
We would like to propose call these SU(2) couplings to hyperspin, since they exhibit spin-like SU(2) behaviours similar to spin, yet, it is different from spin.
For elementary particles like quarks, states with different charged particles cannot be realised at all, due to the super-selection rule, such that composite particles like a neutron and a proton cannot be their superposition state \cite{Georgi99}.
However, for coherent photons, we can consider a superposition state among different topologically charged states, such that we can mix no-vortex state and vortexed states at an arbitrary ratio in amplitudes with a certain definite phase.
Topologically, vortex is well known to be equivalent to a shape of doughnut, which cannot be continuously changed to be a ball.
Our challenge could be considered to realise a superposition state between a doughnut and a ball, which is impossible classically, while we would have a chance, since photons are elementary particles with a wave character to allow a superposition state of orthogonal states.

Here, we consider the hyperspin coupling, which means that we will explore mixing between vortexed states and no-vortex state.
In order to achieve it, an easiest option is to follow the previous approach of the SU(2) state between left and right vortices.
We just need to change from $\hat{\lambda}_2/2=\hat{e}_2^{(t)}$ and $\hat{\lambda}_3/2=\hat{e}_3^{(t)}$ to $\hat{\lambda}_5/2=\hat{e}_2^{(v)}$ and $\hat{e}_3^{(v)}$, respectively, and we define
\begin{eqnarray}
\hat{\mathcal D}_2^{(v)} \left( \theta_y \right)
&=&\exp \left( -i \hat{e}_2^{(v)} \theta_y \right)
\\
&=&
\begin{pmatrix}
\cos \left( \frac{\theta_y}{2} \right) & 0 & - \sin \left( \frac{\theta_y}{2} \right)  \\
0 & 0 & 0\\
\sin \left( \frac{\theta_y}{2} \right) & 0 & \cos \left( \frac{\theta_y}{2} \right)  
\end{pmatrix}
,
\end{eqnarray}
and
\begin{eqnarray}
\hat{\mathcal D}_3^{(v)} \left( \phi_y \right)
&=&\exp \left( -i \hat{e}_3^{(v)} \phi_y \right)
\\
&=&
\begin{pmatrix}
\exp \left( -i\frac{\phi_y}{2} \right) & 0 & 0 \\
0 & 0 & 0 \\
0 & 0 & \exp \left( i\frac{\phi_y}{2} \right)   
\end{pmatrix}
\end{eqnarray}
and we obtain a general SU(3) state, 
\begin{eqnarray}
&&
|
\theta_l,
\phi_l ;
\theta_y,
\phi_y
\rangle
\nonumber \\
&&=
\hat{\mathcal D}_3       \left( \phi_l \right)
\hat{\mathcal D}_2       \left( \theta_l \right)
\hat{\mathcal D}_3^{(v)} \left( \phi_y \right)
\hat{\mathcal D}_2^{(v)} \left( \theta_y \right)
|\psi_1 \rangle
\nonumber
\\
&&=
\left (
  \begin{array}{c}
    {\rm e}^{-i\frac{\phi_y}{2}}
    {\rm e}^{-i\frac{\phi_l}{2}}  
    \cos \left( \frac{\theta_l}{2} \right)    
    \cos \left( \frac{\theta_y}{2} \right)     \\
    {\rm e}^{-i\frac{\phi_y}{2}}
    {\rm e}^{+i\frac{\phi_l}{2}}
    \sin \left( \frac{\theta_l}{2} \right)  
    \cos \left( \frac{\theta_y}{2} \right)     \\
    {\rm e}^{+i\frac{\phi_y}{2}}
    \sin \left( \frac{\theta_y}{2} \right)     
  \end{array}
\right)
.
\nonumber 
\end{eqnarray}

Finally, we can calculate the expectation values for all generators of the $\mathfrak{su}(3)$ Lie algebra, which becomes a vector in an 8-dimensional space, given by 
\begin{eqnarray}
\overrightarrow{\lambda}
&=&
\left (
  \begin{array}{c}
  \lambda_1 \\
  \lambda_2 \\
  \lambda_3 \\
  \lambda_4 \\
  \lambda_5 \\
  \lambda_6 \\
  \lambda_7 \\
  \lambda_8  
    \end{array}
\right)
\nonumber
\\
&=&
\left (
  \begin{array}{c}
    \sin \left( \theta_l \right)    
    \cos \left( \phi_l \right)    
    \cos^2 \left( \frac{\theta_y}{2} \right)    \\
    \sin \left( \theta_l \right)    
    \sin \left( \phi_l \right)    
    \cos^2 \left( \frac{\theta_y}{2} \right)    \\
    \cos \left( \theta_l \right)    
    \cos^2 \left( \frac{\theta_y}{2} \right)    \\
\cos \left( \phi_y + \frac{\phi_l}{2} \right)    
\sin  \left( \theta_y \right) 
\cos \left( \frac{\theta_l}{2} \right)    \\
\sin \left( \phi_y + \frac{\phi_l}{2} \right)    
\sin  \left( \theta_y \right) 
\cos \left( \frac{\theta_l}{2} \right)    \\
\cos \left( \phi_y - \frac{\phi_l}{2} \right)    
\sin  \left( \theta_y \right) 
\sin \left( \frac{\theta_l}{2} \right)    \\
\sin \left( \phi_y - \frac{\phi_l}{2} \right)    
\sin  \left( \theta_y \right) 
\sin \left( \frac{\theta_l}{2} \right)    \\
-
\frac{\sqrt{3}}{6}
+
\frac{\sqrt{3}}{2}
    \cos \left( \theta_y \right)    
  \end{array}
\right)
.
\nonumber 
\end{eqnarray}
An arbitrary state of SU(3) is characterised by this vector, similar to the Stokes parameters \cite{Stokes51} on the Poincar\'e sphere \cite{Poincare92}.
The higher dimensional vector of $\overrightarrow{\lambda}$ satisfies the norm conservation 
\begin{eqnarray}
\sum_{i=1}^{8}
\lambda_i^2
=
\frac{4}{3}
,
\end{eqnarray}
upon rotations in 8-dimensional space, which is actually guaranteed from the constant Casimir operator of $\hat{C}_1=4/3$, as we have seen above.
Therefore, an SU(3) state is represented as a point on the hypersphere with the radius of 
\begin{eqnarray}
\sqrt{
\sum_{i=1}^{8}
\lambda_i^2
}
=
\frac{2}{\sqrt{3}}
.
\end{eqnarray}
We would like to propose this hypersphere as {\it the Gell-Mann hypersphere}, named after Gell-Mann, who found the SU(3) symmetry of baryons and mesons, leading to the discovery of quarks \cite{Gell-Mann61,Gell-Mann64,Ne'eman61}.
In fact, we have {\it the eightfold way} \cite{Gell-Mann61,Gell-Mann64,Ne'eman61} to allow the SU(3) superposition state by changing the amplitudes and the phases of the wavefunction.
We can attribute colour charge of red, green, and blue to 3 fundamental states of $|\psi_1 \rangle$, $|\psi_2 \rangle$, and $|\psi_3 \rangle$, similar to QCD \cite{Gell-Mann61,Gell-Mann64,Ne'eman61,Pfeifer03,Weinberg05}.
In QCD for quarks, only certain sets of multiplets, such as baryons and mesons, are observed as stable bound states of quarks, due to the spontaneous symmetry breaking of the universe \cite{Nambu59,Goldstone62,Higgs64,Anderson58,Schrieffer71}. 
In our photonic QCD, on the other hand, we can discuss an arbitrary superposition state by mixing 3 orthogonal states of left- and right-vorticies and no-vortexed rays.
Therefore, we can discuss the SU(3) state before the symmetry broken, or in other words, the symmetry can be recovered without injecting additional energies to the system, similar to the Nambu-Goldstone bosons \cite{Nambu59,Goldstone62,Higgs64,Anderson58,Schrieffer71}.
This corresponds to rotate the hyperspin of $\overrightarrow{\lambda}$ in the 8-dimensional Gell-Mann space by using 8 generators of rotation $\hat{\lambda}_i$ ($i=1,$ $\cdots$, $8$) to change the amplitudes and the phases.
In experiments, this will be achieved by using rotators and phase-shifters of SU(2) \cite{Saito20a,Saito20b,Saito20c,Saito20d,Saito20e,Saito21f,Saito22g,Saito22h,Saito22i}, since we can realise arbitrary rotations of the SU(3) state by using 3 sets of SU(2) rotations, as we have shown above.

Among 8 Gell-Mann parameters, $\hat{\lambda}_i$, 2 of them are especially important, since the $\mathfrak{su}(3)$ algebra is rank of 2.
One of them is $\ell_3=\lambda_3$, which determines the average orbital angular momentum along the direction of propagation, $z$.
The other important parameter is 
\begin{eqnarray}
y_3
=
\frac{1}{\sqrt{3}}
\lambda_8=
-
\frac{1}{6}
+
\frac{1}{2}
\cos  \left( \theta_y \right) , 
\end{eqnarray}
which determine the average hypercharge.
We confirm the expected maximum and minimum of hypercharge, as $\max \left( y_3 \right)=1/3$ and $\min \left( y_3 \right)=-2/3$.
Hyperchage is simply converted to the topological charge, $q_{\rm t}=y_3+2/3$, and we confirm as $\max \left( q_{\rm t} \right)=1$ and $\min \left( q_{\rm t} \right)=0$, as expected for vorticies and no-vortexed state, respectively.

The Gell-Mann parameters are composed of 8 real parameters and the vector, $\overrightarrow{\lambda}$ has a unit length, $|\overrightarrow{\lambda}|=1$.
Therefore, the rotation of the vector $\overrightarrow{\lambda}$ is achieved by the Special-Orthogonal group of 8 dimensions, SO(8).
The corresponding generators of $\mathfrak{so}(8)$ Lie algebra, are adjoint representations of the $\mathfrak{su}(3)$ generators ($\hat{\lambda}_i$, $i=1,\cdots, 8$) which actually become the structure constants of $C_{ijk}$ ($i,j,k=1,\cdots, 8$).

\subsection{Hyperspin with left/right vortex}

The Gell-Mann hypersphere contains all practical information on the SU(3) states, in terms of amplitudes and phases.
Unfortunately, it is impossible to recognise the 8-dimensional hypersphere for us to see in the world of 3-dimensional space and time.
In the previous sub-section, we have seen the coupling between left and right vortices could be represented by the Poincar\'e sphere for the vortexed photons \cite{Padgett99,Saito21f}, which corresponds to visualise the coupling, controlled by the $\mathfrak{su}(2)$ generators of $\hat{e}_1^{(t)}$, $\hat{e}_2^{(t)}$, and $\hat{e}_3^{(t)}$.
Here, we consider to see another $\mathfrak{su}(2)$ generators and discuss how hyperspin is represented in a similar way to the Poincar\'e sphere.

First, we consider the coupling between the left-vortexed state and no-vortex state.
This corresponds to take the limit $(\theta_l, \phi_l ) \rightarrow (0,0)$, and the Gell-Mann parameters become 
\begin{eqnarray}
\overrightarrow{\lambda}
=
\left (
  \begin{array}{c}
0 \\
0 \\
\frac{1}{2}
\left(
1+ \cos (\theta_y)
\right)\\
    \sin \left( \theta_y \right)    
    \cos \left( \phi_y \right)    \\
    \sin \left( \theta_y \right)    
    \sin \left( \phi_y \right)    \\
0\\
0\\
-
\frac{\sqrt{3}}{6}
+
\frac{\sqrt{3}}{2}
    \cos \left( \theta_y \right)    
  \end{array}
\right)
.
\end{eqnarray}
In this case, the parameter $\lambda_3$ can take a value between 1 and 0, since the left vortex has topological charge of 1 duet to $\hat{\lambda_3} |{\rm L}\rangle = |{\rm L}\rangle $, while the no-vortex state ($|{\rm O}\rangle$), does not have topological charge, as $\hat{\lambda_3} |{\rm O}\rangle = 0$.
The superposition state is characterised non-zero average of $\lambda_3$, and if the amount of the right-vortex component is less than that of the left-vortex, $\lambda_3$ becomes positive.
This corresponds to the net left-circulation of orbital angular momentum.
The other Gell-Mann parameters are given by $\theta_y$ and $\phi_y$.
In the limit of the zero right-vortex component, it is convenient to consider the average of the $\mathfrak{su}(2)$ generating vector, 
\begin{eqnarray}
{\bf v}
&=& (v_1,v_2,v_3)
=\langle \hat{\bf v} \rangle
\nonumber
\\
&=&
(\lambda_4/2,\lambda_5/2,v_3)
\nonumber
\\
&=&
\frac{1}{2}
\left (
  \begin{array}{c}
\sin \left( \theta_y \right)  \cos \left( \phi_y \right)    \\
\sin \left( \theta_y \right)  \sin \left( \phi_y \right)     \\ 
\cos \left( \theta_y \right)    
  \end{array}
\right ),
\end{eqnarray}
which corresponds to introduce $v_3=(\lambda_3+\sqrt{3}\lambda_8)/4$, instead of $\lambda_8$ or $t_{\rm q}$, since only 2 parameters are independent among $(t_3,u_3,v_3)$ due to the rank-2 character of the $\mathfrak{su}(3)$ algebra.

Similarly, we also consider to check the coupling between the right-vortex state and the no-vortex state, which corresponds to take the limit of $(\theta_l, \phi_l )\rightarrow (\pi,0)$, and we obtain the Gell-Mann parameters, 
\begin{eqnarray}
\overrightarrow{\lambda}
=
\left (
  \begin{array}{c}
0 \\
0 \\
-
\frac{1}{2}
\left(
1+ \cos (\theta_y)
\right)\\
0 \\
0 \\
    \sin \left( \theta_y \right)    
    \cos \left( \phi_y \right)    \\
    \sin \left( \theta_y \right)    
    \sin \left( \phi_y \right)    \\
-
\frac{\sqrt{3}}{6}
+
\frac{\sqrt{3}}{2}
    \cos \left( \theta_y \right)    
  \end{array}
\right)
.
\end{eqnarray}
In this case, the sign of $\lambda_3$ changed, compared with the coupling to the left-vortex, since we assigned negative sign for $\hat{\lambda_3} |{\rm R}\rangle = - |{\rm R}\rangle $ to the right-vortex, seen from the observer side against the light, coming to the detector \cite{Saito20a,Saito21f}.
Therefore, $\lambda_3$ can take the value between -1 and 0, which corresponds to the average right-circulation of orbital angular momentum.
For the right circulation, it is useful to calculate the average of the $\mathfrak{su}(2)$ generating vector, $ \hat{\bf u}$ as
\begin{eqnarray}
{\bf u}
&=& (u_1,u_2,u_3)
=\langle \hat{\bf u} \rangle
\nonumber
\\
&=&
(\lambda_6/2,\lambda_7/2,u_3)
\nonumber
\\
&=&
\frac{1}{2}
\left (
  \begin{array}{c}
\sin \left( \theta_y \right)  \cos \left( \phi_y \right)    \\
\sin \left( \theta_y \right)  \sin \left( \phi_y \right)     \\ 
\cos \left( \theta_y \right)    
  \end{array}
\right ),
\end{eqnarray}
which becomes the same formula for ${\bf v}$, when only 1 chirality (i.e., left or right vortex) is involved.
In fact, the parameters $\theta_y$ and $\phi_y$ account for the relative phase and the amplitudes between the state with $|m|=1$ and the state with $m=0$ without including the difference in chiralities.
Here, $u_3$ is introduced by $u_3=(-\lambda_3+\sqrt{3}\lambda_8)/4$, and it satisfies the conservation law of $v_3-u_3=t_3$.

Consequently, we have obtained 3 vectors, $({\bf t},{\bf u},{\bf v})$, where ${\bf t}=(\ell_1,\ell_2,\ell_3)/2$ is coming from the average orbital angular momentum.
Each vector of ${\bf t}$, ${\bf u}$, or ${\bf v}$ is 3-dimensional, such that they are represented by the Poincar\'e spheres.
However, care must be taken in the radiuses of the Poincar\'e spheres, since they depend on the relative amplitudes, determined by $\theta_{l}$ and $\theta_y$.
This comes from the mutual dependence among 3 sets of the $\mathfrak{su}(2)$ algebra, since the $\mathfrak{su}(3)$ algebra does not contain the non-trivial invariant group, as we confirmed above.
As a result, we obtained 3 mutually dependent spheres, which have $3\times 3=9$ parameters, with 1 identity of $v_3=u_3+t_3$.
For visualisation purposes, the 3 Poincar\'e spheres with variable radiuses might be practically more useful for humans, living in 3 spatial dimensions, rather than 8-dimensional Gell-Mann hypersphere of the constant radius of $2/\sqrt{3}$, whose surface is equivalent to 7-dimensional spherical surface of ${\mathbb S}^7$, given by real numbers, with fixed radius in 8-dimensional space.

\subsection{Hyperspin embedded in SO(6)}

Gell-Mann parameters in SO(8) are useful to understand the coupling between $|{\rm L}\rangle$, $|{\rm R}\rangle$, and $|{\rm O}\rangle$.
However, we can easily recognise that the generators of  $\mathfrak{su}(3)$ cannot span the whole hypersurface of SO(8).
For example, parameters $\lambda_i$ ($i=1,\cdots,7$), except for $\lambda_8$, cannot take values above 1 nor below -1, while the radius of $2/\sqrt{3}$ is larger than 1.
This clearly shows a point like $(2/\sqrt{3},0,\cdots, 0)$ cannot be covered at all, such that SO(8) is much larger than parameter space required to represent the photonic states, composed of 3 orthogonal states.

Then, let's consider the number of freedom, required for mixing $|{\rm L}\rangle$, $|{\rm R}\rangle$, and $|{\rm O}\rangle$.
In general, we should consider variable density of photons, since the radius of the Poincar\'e sphere depends on the output power of the ray \cite{Stokes51,Poincare92,Jones41,Fano54,Baym69,Sakurai14,Max99,Jackson99,Yariv97,Gil16,Goldstein11,
Hecht17,Pedrotti07,Saito20a,Saito20b,Saito20c,Saito20d,Saito21f,Saito22g,Saito22h}.
Then, photons in coherent state are represented by 1 complex number per orthogonal degree of freedom for the component of the wavefunction.
We are considering for fixed polarisation state, while we have 3 orthogonal states for vorticies, and therefore, we have 6 degrees of freedom (Table III).


\begin{table}
\caption{\label{Table-III}
Degrees of freedom for photons with 3 orthogonal states.}
\begin{ruledtabular}
\begin{tabular}{clcc}
&Variables & Degrees of freedom & \\
\colrule
&Power density of the ray & 1 & \\
&Global phase & 1 & \\
&Orbital angular momentum & 2 & \\
&Hyperspin & 2 &
\end{tabular}
\end{ruledtabular}
\end{table}

These 6 degrees of freedom are attributed to corresponding physical parameters (Table III).
1 degree is assigned to the power density of the ray, and another is used for the global U(1) phase, which will not play a role for expectation values of the Gell-Mann hypersphere.
2 degrees of freedom are required for describing the superposition state for orbital angular momentum, which can be shown in the Poincar\'e sphere with variable radius (Fig. 2 (a)).
Therefore, the rest of the remaining 2 parameters should be assigned to hyperspin to account for the mixing of $|{\rm L}\rangle$ and/or $|{\rm R}\rangle$ with $|{\rm O}\rangle$.
This picture is consistent with the wavefunction of $|\theta_l,\phi_l ;\theta_y,\phi_y \rangle$, where $\theta_y$ and $\phi_y$ account for hyperspin.
On the other hand, we used 8 Gell-Mann parameters for describing the superposition state from the expectation values.
All 8 parameters are required to understand the full rotational ways on the Gell-Mann hypersphere, however, less parameters are required to scan the full wavefunction over the expected Hilbert space of ${\mathbb S}^5$.
Here, we try to reduce the number of Gell-Mann parameters to embed hyperspin in SO(6).
The goal is to represent hyperspin
\begin{eqnarray}
y_1
&=&
\frac{1}{2}
\sin  \left( \theta_y \right) 
\cos  \left( \phi_y \right) 
\\
y_2
&=&
\frac{1}{2}
\sin  \left( \theta_y \right) 
\sin  \left( \phi_y \right) 
\\
y_3
&=&
-
\frac{1}{6}
+
\frac{1}{2}
\cos  \left( \theta_y \right) 
,
\end{eqnarray}
as shown on the Poincar\'e sphere (Fig. 2 (b)), which should be enough for showing $\theta_y$ and $\phi_y$, topologically. 

A hint is found in Gell-Mann parameters, $\lambda_4$, $\cdots$, $\lambda_7$, which keep the magnitude, 
\begin{eqnarray}
\left(
\lambda_4
\right)^2
+
\left(
\lambda_5
\right)^2
+
\left(
\lambda_6
\right)^2
+
\left(
\lambda_7
\right)^2
=
\sin^2
\left(
\theta_y
\right)
\end{eqnarray}
upon changing other parameters of $\theta_l$, $\phi_l$, and $\phi_y$.
Therefore, we have a chance to eliminate $\lambda_6$ and $\lambda_7$ by the renormalising the operators for the $\mathfrak{su}(3)$ algebra.

By inspecting $\lambda_4$, $\cdots$, $\lambda_7$, we realise the phases of $\phi_y$ and $\phi_l$ are coupled in a mixed form.
If we could convert $\phi_y \pm \phi_l/2 \rightarrow \phi_y$, the rest of parameters are easily converted upon rotations.
This could be achieved, if we remember the rotation matrices of 
\begin{eqnarray}
{\mathcal R}
\left( \theta \right)    
=
\left (
  \begin{array}{cc}
\cos \theta  & - \sin \theta 
\\
\sin \theta & \cos \theta \\
  \end{array}
\right)
\end{eqnarray}
form a group to satisfy the associative requirement
\begin{eqnarray}
{\mathcal R} \left( \phi_y + \frac{\phi_l}{2} \right)    
=
{\mathcal R} \left( \frac{\phi_l}{2} \right)    
{\mathcal R} \left( \phi_y  \right)    
.
\end{eqnarray}
Then, we obtain
\begin{eqnarray}
\left (
  \begin{array}{c}
\cos \left( \phi_y + \frac{\phi_l}{2} \right)    
\\
\sin \left( \phi_y + \frac{\phi_l}{2} \right)    
\\
  \end{array}
\right)
=
{\mathcal R}
\left( \frac{\phi_l}{2} \right)    
\left (
  \begin{array}{c}
\cos \left( \phi_y  \right)    
\\
\sin \left( \phi_y  \right)    
\\
  \end{array}
\right)
,
\nonumber
\end{eqnarray}
whose reverse relationship becomes
\begin{eqnarray}
\left (
  \begin{array}{c}
\cos \left( \phi_y  \right)    
\\
\sin \left( \phi_y  \right)    
\\
  \end{array}
\right)
=
{\mathcal R}
\left( -\frac{\phi_l}{2} \right)    
\left (
  \begin{array}{c}
\cos \left( \phi_y + \frac{\phi_l}{2} \right)    
\\
\sin \left( \phi_y + \frac{\phi_l}{2} \right)    
\\
  \end{array}
\right)
.
\nonumber
\end{eqnarray}
By using these formulas, we define
\begin{eqnarray}
\left (
  \begin{array}{c}
\lambda_4^{\prime}
\\
\lambda_5^{\prime}
\\
  \end{array}
\right)
&=&
{\mathcal R}
\left( -\frac{\phi_l}{2} \right)    
\left (
  \begin{array}{c}
\lambda_4
\\
\lambda_5
  \end{array}
\right)
\nonumber
\\
&=&
\left (
  \begin{array}{cc}
\cos \left(\frac{\phi_l}{2} \right)    
&
\sin \left( \frac{\phi_l}{2} \right)    
\\
-\sin \left( \frac{\phi_l}{2} \right)    
&
\cos \left(\frac{\phi_l}{2} \right)    
  \end{array}
\right)
\left (
  \begin{array}{c}
\lambda_4
\\
\lambda_5
  \end{array}
\right)
\nonumber
\\
&=&
\left (
  \begin{array}{c}
\cos \left(\phi_y \right)    
\sin \left(\theta_y \right)    
\cos \left(\frac{\theta_l}{2} \right)    
\\
\sin \left(\phi_y \right)    
\sin \left(\theta_y \right)    
\cos \left(\frac{\theta_l}{2} \right)    
  \end{array}
\right)
\nonumber
\end{eqnarray}
and
\begin{eqnarray}
\left (
  \begin{array}{c}
\lambda_6^{\prime}
\\
\lambda_7^{\prime}
\\
  \end{array}
\right)
&=&
{\mathcal R}
\left( \frac{\phi_l}{2} \right)    
\left (
  \begin{array}{c}
\lambda_6
\\
\lambda_7
\\
  \end{array}
\right)
\nonumber
\\
&=&
\left (
  \begin{array}{cc}
\cos \left(\frac{\phi_l}{2} \right)    
&
-\sin \left( \frac{\phi_l}{2} \right)    
\\
\sin \left( \frac{\phi_l}{2} \right)    
&
\cos \left(\frac{\phi_l}{2} \right)    
\\
  \end{array}
\right)
\left (
  \begin{array}{c}
\lambda_6
\\
\lambda_7
\\
  \end{array}
\right)
\nonumber
\\
&=&
\left (
  \begin{array}{c}
\cos \left(\phi_y \right)    
\sin \left(\theta_y \right)    
\sin \left(\frac{\theta_l}{2} \right)    
\\
\sin \left(\phi_y \right)    
\sin \left(\theta_y \right)    
\sin \left(\frac{\theta_l}{2} \right)    
  \end{array}
\right)
.
\nonumber
\end{eqnarray}
Then, we could successfully convert $\phi_y \pm \phi_l/2 \rightarrow \phi_y$, as intended.
Finally, we can rotate between $\lambda_4$ and $\lambda_6$ to eliminate $\lambda_6$ by defining \begin{eqnarray}
\left (
  \begin{array}{c}
\lambda_4^{\prime \prime}
\\
\lambda_6^{\prime \prime}
\\
  \end{array}
\right)
&=&
{\mathcal R}
\left( -\frac{\theta_l}{2} \right)    
\left (
  \begin{array}{c}
\lambda_4^{\prime}
\\
\lambda_6^{\prime}
\\
  \end{array}
\right)
\\
&=&
\left (
  \begin{array}{c}
\cos \left(\phi_y \right)    
\sin \left(\theta_y \right)    
\\
0\\
  \end{array}
\right)
.
\end{eqnarray}
Similarly, we define

\begin{eqnarray}
\left (
  \begin{array}{c}
\lambda_5^{\prime \prime}
\\
\lambda_7^{\prime \prime}
\\
  \end{array}
\right)
&=&
{\mathcal R}
\left( -\frac{\theta_l}{2} \right)    
\left (
  \begin{array}{c}
\lambda_5^{\prime}
\\
\lambda_7^{\prime}
\\
  \end{array}
\right)
\\
&=&
\left (
  \begin{array}{c}
\sin \left(\phi_y \right)    
\sin \left(\theta_y \right)    
\\
0\\
  \end{array}
\right)
.
\end{eqnarray}

In order to obtain these expectation values for Gell-Mann parameters, we should renormalise the original $\mathfrak{su}(3)$ basis operators to define
\begin{eqnarray}
&&
\hat{\lambda}_4^{\prime \prime}
=
\cos \left(\frac{\theta_l}{2} \right)    
\hat{\lambda}_4^{\prime }
+
\sin \left(\frac{\theta_l}{2} \right)    
\hat{\lambda}_6^{\prime }
\nonumber
\\
&&=
\cos \left(\frac{\theta_l}{2} \right)    
\left(
  \cos \left(\frac{\phi_l}{2} \right)    
  \hat{\lambda}_4
+
  \sin \left(\frac{\phi_l}{2} \right)    
  \hat{\lambda}_5
\right)
\nonumber
\\
&& \ \ 
+
\sin \left(\frac{\theta_l}{2} \right)    
\left(
  \cos \left(\frac{\phi_l}{2} \right)    
  \hat{\lambda}_6
-
  \sin \left(\frac{\phi_l}{2} \right)    
  \hat{\lambda}_7
\right)
\nonumber
\\
&&=
\left (
  \begin{array}{ccc}
   0   &   0   &    {\rm e}^{-i\frac{\phi_l}{2}} \cos \left(\frac{\theta_l}{2} \right) \\
   0   &   0   &    {\rm e}^{i\frac{\phi_l}{2}}  \sin \left(\frac{\theta_l}{2} \right) \\
   {\rm e}^{i\frac{\phi_l}{2}} \cos \left(\frac{\theta_l}{2} \right)      &   {\rm e}^{-i\frac{\phi_l}{2}} \sin \left(\frac{\theta_l}{2} \right)   &    0 
  \end{array}
\right)
\nonumber
\end{eqnarray}
and
\begin{eqnarray}
&&\hat{\lambda}_5^{\prime \prime}
=
\cos \left(\frac{\theta_l}{2} \right)    
\hat{\lambda}_5^{\prime }
+
\sin \left(\frac{\theta_l}{2} \right)    
\hat{\lambda}_7^{\prime }
\nonumber
\\
&&=
\cos \left(\frac{\theta_l}{2} \right)    
\left(
  -\sin \left(\frac{\phi_l}{2} \right)    
  \hat{\lambda}_4
+
  \cos \left(\frac{\phi_l}{2} \right)    
  \hat{\lambda}_5
\right)
\nonumber
\\
&& \ \ 
+
\sin \left(\frac{\theta_l}{2} \right)    
\left(
  \sin \left(\frac{\phi_l}{2} \right)    
  \hat{\lambda}_6
+
  \cos \left(\frac{\phi_l}{2} \right)    
  \hat{\lambda}_7
\right)
\nonumber
\\
&&=
\left (
  \begin{array}{ccc}
   0   &   0   &    -i{\rm e}^{-i\frac{\phi_l}{2}} \cos \left(\frac{\theta_l}{2} \right) \\
   0   &   0   &    -i{\rm e}^{i\frac{\phi_l}{2}}  \sin \left(\frac{\theta_l}{2} \right) \\
   i{\rm e}^{i\frac{\phi_l}{2}} \cos \left(\frac{\theta_l}{2} \right)      &   i{\rm e}^{-i\frac{\phi_l}{2}} \sin \left(\frac{\theta_l}{2} \right)   &    0 
  \end{array}
\right)
.
\nonumber 
\end{eqnarray}
If we use these operators, the Gell-Mann parameters become
\begin{eqnarray}
\overrightarrow{\lambda}^{\prime \prime}
=
\left (
  \begin{array}{c}
    \sin \left( \theta_l \right)    
    \cos \left( \phi_l \right)    
    \cos^2 \left( \frac{\theta_y}{2} \right)    \\
    \sin \left( \theta_l \right)    
    \sin \left( \phi_l \right)    
    \cos^2 \left( \frac{\theta_y}{2} \right)    \\
    \cos \left( \theta_l \right)    
    \cos^2 \left( \frac{\theta_y}{2} \right)    \\
\cos \left( \phi_y  \right)    
\sin  \left( \theta_y \right)    \\
\sin \left( \phi_y  \right)    
\sin  \left( \theta_y \right)    \\
0   \\
0   \\
-
\frac{\sqrt{3}}{6}
+
\frac{\sqrt{3}}{2}
    \cos \left( \theta_y \right)    
  \end{array}
\right)
,
\end{eqnarray}
such that we could successfully remove $\lambda_6$ and $\lambda_7$.
These parameters are equivalent to use photonic orbital angular momentum
\begin{eqnarray}
\overrightarrow{\ell}
=
\left (
  \begin{array}{c}
    \sin \left( \theta_l \right)    
    \cos \left( \phi_l \right)    
    \cos^2 \left( \frac{\theta_y}{2} \right)    \\
    \sin \left( \theta_l \right)    
    \sin \left( \phi_l \right)    
    \cos^2 \left( \frac{\theta_y}{2} \right)    \\
    \cos \left( \theta_l \right)    
    \cos^2 \left( \frac{\theta_y}{2} \right)    
  \end{array}
\right)
\end{eqnarray}
and hyperspin
\begin{eqnarray}
\overrightarrow{y}
=
\left (
  \begin{array}{c}
\frac{1}{2}
\sin  \left( \theta_y \right) 
\cos  \left( \phi_y \right)   \\
\frac{1}{2}
\sin  \left( \theta_y \right) 
\sin  \left( \phi_y \right) 
      \\
-
\frac{1}{6}
+
\frac{1}{2}
\cos  \left( \theta_y \right) 
  \end{array}
\right)
,
\end{eqnarray}
which can be shown on 2 Poincar\'e spheres with the radiuses of $ \cos^2 (\theta_y/2) $ and $1/2$, respectively, instead of original 3 spheres.
This is consistent with 4 degrees of freedom for orbital angular momentum and hyperspin, as confirmed before (Table III), and it is also expected from the rank-2 nature of $\mathfrak{su}(3)$ algebra, which requires only 2 sets of $\mathfrak{su}(3)$ among 3 sets of $({\bf \hat{t}},{\bf \hat{u}},{\bf \hat{v}})$.
In practice, we do not know the angles of $\theta_l$ and $\phi_l$, {\it a priori}, such that the angles are obtained from expectation values or experimental results.

Finally, we could successfully embed Gell-Mann parameters in SO(6) to renormalise
\begin{eqnarray}
\overrightarrow{S}
&=&
\left (
  \begin{array}{c}
S_1\\
S_2\\
S_3\\
S_4\\
S_5\\
S_6
  \end{array}
\right)
=
\left (
  \begin{array}{c}
\lambda_1\\
\lambda_2\\
\lambda_3\\
\lambda_4^{\prime \prime}\\
\lambda_5^{\prime \prime}\\
\lambda_8
  \end{array}
\right)
\nonumber
\\
&=&
\left (
  \begin{array}{c}
    \sin \left( \theta_l \right)    
    \cos \left( \phi_l \right)    
    \cos^2 \left( \frac{\theta_y}{2} \right)    \\
    \sin \left( \theta_l \right)    
    \sin \left( \phi_l \right)    
    \cos^2 \left( \frac{\theta_y}{2} \right)    \\
    \cos \left( \theta_l \right)    
    \cos^2 \left( \frac{\theta_y}{2} \right)    \\
\cos \left( \phi_y  \right)    
\sin  \left( \theta_y \right)    \\
\sin \left( \phi_y  \right)    
\sin  \left( \theta_y \right)    \\
-
\frac{\sqrt{3}}{6}
+
\frac{\sqrt{3}}{2}
    \cos \left( \theta_y \right)    
  \end{array}
\right)
,
\end{eqnarray}
which satisfies the conservation law of the norm, 
\begin{eqnarray}
\sum_{i=1}^6
\left( 
S_i
\right)^2
=
\frac{4}{3}
,
\end{eqnarray}
which was required from the constant Casimir operator of $\hat{C}_1$.

\begin{figure}[h]
\begin{center}
\includegraphics[width=8cm]{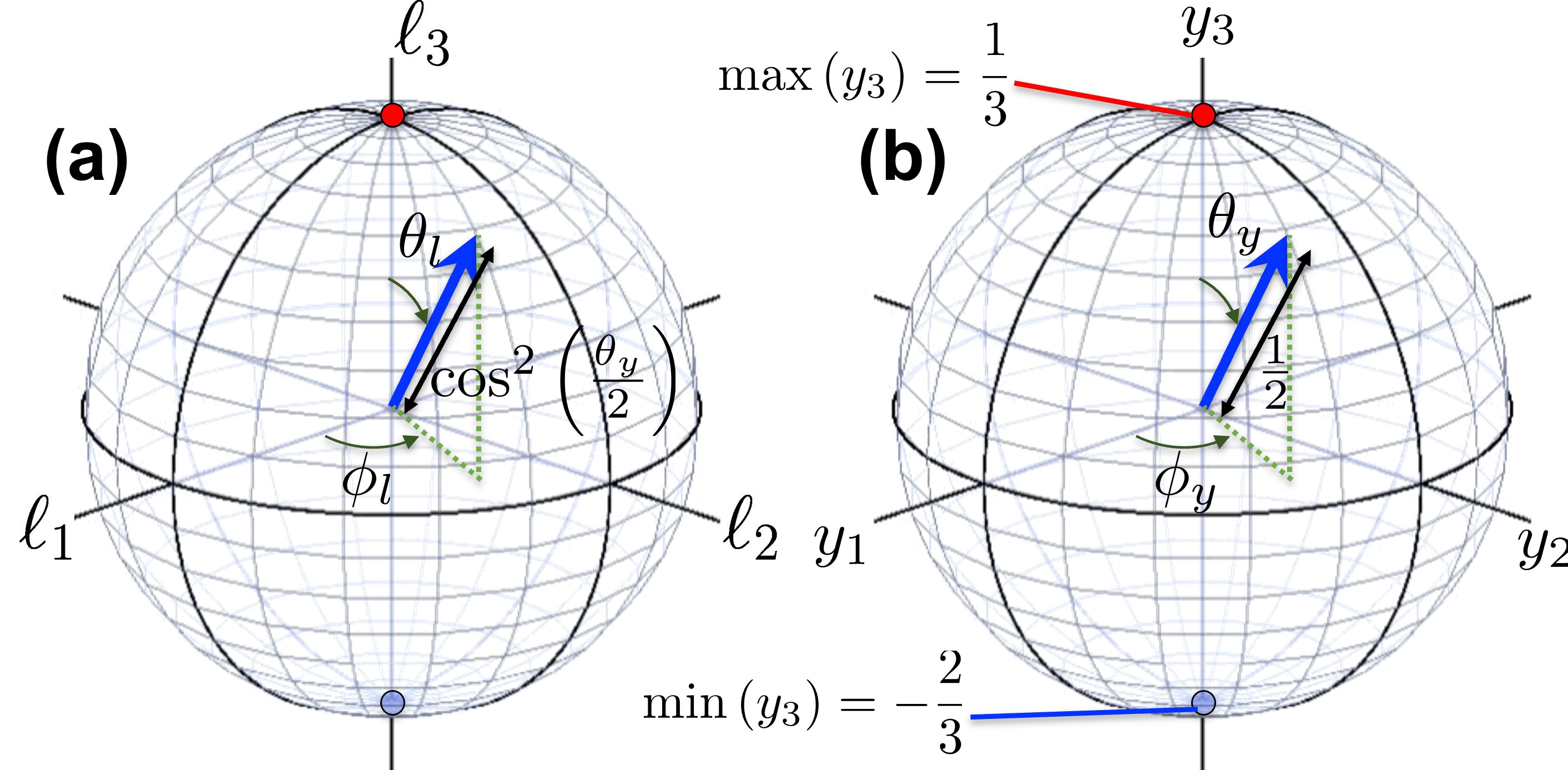}
\caption{
Renormalisation of Gell-Mann parameters.
8-dimensional Gell-Mann hypersphere could be reduced to 2 Poincar\'e spheres for (a) photonic orbital angular momentum and (b) hyperspin.
The radius of the Poincar\'e sphere for orbital angular momentum is $ \cos^2 ( \theta_y/2) $, while it is $1/2$ for hyperspin.
The maximum and minimum of $y_3$ correspond to hypercharge of 1/3 and -2/3, which are equivalent to topological charge of 1 (pure vortex of $|{\rm L \rangle}$ or $|{\rm R}\rangle$) and 0 (no vortex, $|{\rm O}\rangle$), respectively.
}
\end{center}
\end{figure}

\subsection{Alternative Coherent States}
We have used $\hat{\mathcal D}_2^{(v)} \left( \theta_y \right)$ and $\hat{\mathcal D}_3^{(v)} \left( \phi_y \right)$ to define an arbitrary state, but we can define alternative coherent state, using original $\mathfrak{su}(3)$ bases by
\begin{eqnarray}
&&\hat{\mathcal D}_8 \left( \phi_y \right)
=\exp \left( -i \hat{\lambda}_8 \frac{\phi_y}{2} \right)
\nonumber
\\
&&=
\begin{pmatrix}
\exp \left( -i \frac{1}{\sqrt{3}} \frac{\phi_y}{2} \right) & 0 & 0 \\
0 & \exp \left( -i \frac{1}{\sqrt{3}} \frac{\phi_y}{2} \right)  & 0 \\
0 & 0 & \exp \left( i \frac{2}{\sqrt{3}} \frac{\phi_y}{2} \right)   
\end{pmatrix}
\nonumber
\end{eqnarray}
and $\hat{\mathcal D}_5 \left( \theta_y \right)=\hat{\mathcal D}_2^{(v)} \left( \theta_y \right)$, as
\begin{eqnarray}
|
\theta_l,
\phi_l ;
\theta_y,
\phi_y
\rangle
&=&
\hat{\mathcal D}_8 \left( \phi_y \right)
\hat{\mathcal D}_3       \left( \phi_l \right)
\hat{\mathcal D}_2       \left( \theta_l \right)
\hat{\mathcal D}_5 \left( \theta_y \right)
|\psi_1 \rangle
\nonumber
\\
&=&
\left (
  \begin{array}{c}
    {\rm e}^{-i\frac{1}{\sqrt{3}}\frac{\phi_y}{2}}
    {\rm e}^{-i\frac{\phi_l}{2}}  
    \cos \left( \frac{\theta_l}{2} \right)    
    \cos \left( \frac{\theta_y}{2} \right)     \\
    {\rm e}^{-i\frac{1}{\sqrt{3}}\frac{\phi_y}{2}}
    {\rm e}^{+i\frac{\phi_l}{2}}
    \sin \left( \frac{\theta_l}{2} \right)  
    \cos \left( \frac{\theta_y}{2} \right)     \\
    {\rm e}^{+i\frac{2}{\sqrt{3}}\frac{\phi_y}{2}}
    \sin \left( \frac{\theta_y}{2} \right)     
  \end{array}
\right)
.
\nonumber
\end{eqnarray}
Using this coherent state, we obtain the Gell-Mann parameters as expectation values, 
\begin{eqnarray}
\overrightarrow{\lambda}
=
\left (
  \begin{array}{c}
    \sin \left( \theta_l \right)    
    \cos \left( \phi_l \right)    
    \cos^2 \left( \frac{\theta_y}{2} \right)    \\
    \sin \left( \theta_l \right)    
    \sin \left( \phi_l \right)    
    \cos^2 \left( \frac{\theta_y}{2} \right)    \\
    \cos \left( \theta_l \right)    
    \cos^2 \left( \frac{\theta_y}{2} \right)    \\
\cos \left( \frac{\sqrt{3}}{2}\phi_y + \frac{\phi_l}{2} \right)    
\sin  \left( \theta_y \right) 
\cos \left( \frac{\theta_l}{2} \right)    \\
\sin \left( \frac{\sqrt{3}}{2}\phi_y + \frac{\phi_l}{2} \right)    
\sin  \left( \theta_y \right) 
\cos \left( \frac{\theta_l}{2} \right)    \\
\cos \left( \frac{\sqrt{3}}{2}\phi_y - \frac{\phi_l}{2} \right)    
\sin  \left( \theta_y \right) 
\sin \left( \frac{\theta_l}{2} \right)    \\
\sin \left( \frac{\sqrt{3}}{2}\phi_y - \frac{\phi_l}{2} \right)    
\sin  \left( \theta_y \right) 
\sin \left( \frac{\theta_l}{2} \right)    \\
-
\frac{\sqrt{3}}{6}
+
\frac{\sqrt{3}}{2}
    \cos \left( \theta_y \right)    
  \end{array}
\right)
.
\nonumber \\
\end{eqnarray}

We will embed Gell-Mann parameters into SO(6) for this coherent state in the same way with the previous subsection.
To achieve such a conversion, we need to transfer $\sqrt{3}\phi_y/2  \pm \phi_l/2 \rightarrow \sqrt{3} \phi_y /2 $, which appeared in $\lambda_4$, $\cdots$, $\lambda_7$, by confirming 
\begin{eqnarray}
\left (
  \begin{array}{c}
\cos \left( \frac{\sqrt{3}}{2} \phi_y + \frac{\phi_l}{2} \right)    
\\
\sin \left( \frac{\sqrt{3}}{2}  \phi_y + \frac{\phi_l}{2} \right)    
\\
  \end{array}
\right)
=
{\mathcal R}
\left( \frac{\phi_l}{2} \right)    
\left (
  \begin{array}{c}
\cos \left(  \frac{\sqrt{3}}{2} \phi_y  \right)    
\\
\sin \left(  \frac{\sqrt{3}}{2} \phi_y  \right)    
\\
  \end{array}
\right)
,
\nonumber 
\end{eqnarray}
whose inverse becomes
\begin{eqnarray}
\left (
  \begin{array}{c}
\cos \left( \frac{\sqrt{3}}{2} \phi_y  \right)    
\\
\sin \left( \frac{\sqrt{3}}{2} \phi_y  \right)    
\\
  \end{array}
\right)
=
{\mathcal R}
\left( -\frac{\phi_l}{2} \right)    
\left (
  \begin{array}{c}
\cos \left( \frac{\sqrt{3}}{2} \phi_y + \frac{\phi_l}{2} \right)    
\\
\sin \left( \frac{\sqrt{3}}{2} \phi_y + \frac{\phi_l}{2} \right)    
\\
  \end{array}
\right)
.
\nonumber
\end{eqnarray}

The rest of the calculations are exactly the same with the previous subsection, and we can use the same renormalised operators of $\hat{\lambda_4}^{\prime \prime}$ $\hat{\lambda_5}^{\prime \prime}$, while we remove $\hat{\lambda_6}^{\prime \prime}=0$ and $\hat{\lambda_7}^{\prime \prime}=0$.
Then, the renormalised Gell-Mann parameters become
\begin{eqnarray}
\overrightarrow{\lambda}^{\prime \prime}
=
\left (
  \begin{array}{c}
    \sin \left( \theta_l \right)    
    \cos \left( \phi_l \right)    
    \cos^2 \left( \frac{\theta_y}{2} \right)    \\
    \sin \left( \theta_l \right)    
    \sin \left( \phi_l \right)    
    \cos^2 \left( \frac{\theta_y}{2} \right)    \\
    \cos \left( \theta_l \right)    
    \cos^2 \left( \frac{\theta_y}{2} \right)    \\
\sin  \left( \theta_y \right)    
\cos \left( \frac{\sqrt{3}}{2} \phi_y  \right)    
\\
\sin  \left( \theta_y \right)    
\sin \left( \frac{\sqrt{3}}{2} \phi_y  \right)    
\\
0   \\
0   \\
-
\frac{\sqrt{3}}{6}
+
\frac{\sqrt{3}}{2}
    \cos \left( \theta_y \right)    
  \end{array}
\right)
,
\end{eqnarray}
which keep $\overrightarrow{\ell}$ unchanged, while hyperspin becomes
\begin{eqnarray}
\overrightarrow{y}
=
\left (
  \begin{array}{c}
\frac{1}{2}
\sin  \left( \theta_y \right) 
\cos  \left( \frac{\sqrt{3}}{2} \phi_y \right)   \\
\frac{1}{2}
\sin  \left( \theta_y \right) 
\sin  \left( \frac{\sqrt{3}}{2}  \phi_y \right) 
      \\
-
\frac{1}{6}
+
\frac{1}{2}
\cos  \left( \theta_y \right) 
  \end{array}
\right)
.
\end{eqnarray}
This just corresponds to change the azimuthal angle, $\phi_y \rightarrow \sqrt{3} \phi_y /2$, in the Poincar\'e sphere of Fig. 2 (b).
Consequently, we could embed Gell-Mann parameters in SO(6) as 
\begin{eqnarray}
\overrightarrow{S}
&=&
\left (
  \begin{array}{c}
S_1\\
S_2\\
S_3\\
S_4\\
S_5\\
S_6
  \end{array}
\right)
=
\left (
  \begin{array}{c}
\lambda_1\\
\lambda_2\\
\lambda_3\\
\lambda_4^{\prime \prime}\\
\lambda_5^{\prime \prime}\\
\lambda_8
  \end{array}
\right)
\nonumber \\
&=&
\left (
  \begin{array}{c}
    \sin \left( \theta_l \right)    
    \cos \left( \phi_l \right)    
    \cos^2 \left( \frac{\theta_y}{2} \right)    \\
    \sin \left( \theta_l \right)    
    \sin \left( \phi_l \right)    
    \cos^2 \left( \frac{\theta_y}{2} \right)    \\
    \cos \left( \theta_l \right)    
    \cos^2 \left( \frac{\theta_y}{2} \right)    \\
\sin  \left( \theta_y \right)    
\cos \left( \frac{\sqrt{3}}{2} \phi_y  \right)   
\\
\sin  \left( \theta_y \right)    
\sin \left(  \frac{\sqrt{3}}{2}  \phi_y  \right)    
\\
-
\frac{\sqrt{3}}{6}
+
\frac{\sqrt{3}}{2}
    \cos \left( \theta_y \right)    
  \end{array}
\right)
,
\end{eqnarray}
which also keep the norm
\begin{eqnarray}
\sum_{i=1}^6
\left( 
S_i
\right)^2
=
\frac{4}{3}
,
\end{eqnarray}
upon arbitrary rotations in 6-dimensional space in SO(6).
In the practical experiments, however, it will be more complex, if we set up a rotator for $\hat{\mathcal D}_8 \left( \phi_y \right)$, since 3 waves are involved rather than 2 waves.
In conventional optical experiments, various splitters and combiners are prepared for 2 waves such that it is much easier to rely on SU(2) rotations, including $\hat{\mathcal D}_3^{(v)} \left( \phi_y \right)$ and $\hat{\mathcal D}_3^{(u)} \left( \phi_y \right)=\exp \left( -i \hat{e}_3^{(u)} \phi_y \right)$, such that we do not have to stick on using original bases of $\hat{\lambda}_i$ for SU(3) states.

\section{Embedding in SO(5)}
For the complete description of the eightfold way to rotate the SU(3) states, Gell-Mann parameters in SO(8) are more useful to understand the differences in phases and amplitudes among $|{\rm L} \rangle$, $|{\rm R}  \rangle$ and $|{\rm O}  \rangle$.
On the other hand, SO(8) is too larger to show the nature of the wavefunction, made of 3 complex numbers (${\mathbb C}^3$) with its norm conserved to cover ${\mathbb S}^5$ in Hilbert space.

We could successfully reduce the dimension of Gell-Mann parameters from SO(8) to SO(6) or $SO(3) \times SO(3)$ to represent SU(3) states, in terms of orbital angular momentum and hyperspin, as expectation values.
On the other hand, we have just 4 parameters ($\theta_l$, $\phi_l$,  $\theta_y$, and $\phi_y$), such that we have a chance to reduce 1 more dimension to represent on ${\mathbb S}^4$ in SO(5).

Before proceeding further, we review the relationship between SU(2) and SO(3) for describing spin states or polarisation states for photons \cite{Stokes51,Poincare92,Jones41,Fano54,Baym69,Sakurai14,Max99,Jackson99,Yariv97,Gil16,Goldstein11,
Hecht17,Pedrotti07,Saito20a,Saito20b,Saito20c,Saito20d}.
For polarisation, we have 2 orthogonal states, such that a ray of coherent photons are described by SU(2) states, which require 2 complex numbers (${\mathbb C}^2$).
The SU(2) wavefunction could be normalised for a fixed power density, such that 1 degree of freedom disappeared and the wavefunction covers ${\mathbb S}^3$ in Hilbert space.
In fact, according to the fundamental theorem of homomorphism \cite{Stubhaug02,Fulton04,Hall03,Pfeifer03,Georgi99,Cisowski22}, 
${\rm SU(2)/SU(1)} \cong {\rm SU(2)/U(1)} \cong {\mathbb S}^3$.
This means that the SU(2) wavefunction is equivalent to ${\mathbb S}^3$, except for the global phase factor of U(1).
On the other hand, it is also well known that ${\rm SU(2)}/{\mathbb S}^0 \cong {\rm SU(2)}/{\mathbb Z}_2 \cong {\rm SO(3)}$, where ${\mathbb S}^0= \{ -1, 1 \}$ and ${\mathbb Z}_2= \{ 0, 1 \}$.
This means that if we neglect the impacts of the global phase factor, such as those expected from the geometrical Pancharatnam-Berry's phases \cite{Pancharatnam56,Berry84} in closed loops, the expectation values of SU(2) states are represented on the sphere, represented by the SO(3) group.
Consequently, the original topology of the wavefunction on ${\mathbb S}^3$ is reduced to the Poincar\'e sphere of ${\mathbb S}^2$ in expectation values.

Similarly, in SU(3), the fundamental theorem of homomorphism \cite{Stubhaug02,Fulton04,Hall03,Pfeifer03,Georgi99} leads ${\rm SU(3)/SU(2)} \cong {\mathbb S}^5$.
This means that we have a freedom of an SU(2) symmetry within SU(3) states, which keeps the states essentially equivalent to ${\mathbb S}^5$, as we have confirmed from the identity of $v_3=u_3+t_3$ to allow arbitrary 2 choices of SU(2) states from 3 sets of SU(2) bases, $(\hat{\bf t},\hat{\bf u},\hat{\bf v})$.
In the similarity with the SU(2) states, one of the degree of freedom in $ {\mathbb S}^5$ would be coming from the global phase, such that we may have a chance to represent the expectation values on ${\mathbb S}^4$ in SO(5).

However, we could not establish a surjective mapping from SO(6) to SO(5) purely upon rotations using our bases of $\mathfrak{su}(3)$, because the expectation values of $\lambda_i$ ($i=1,\cdots, 7$) cannot be larger than 1, while we needed to renormalise $\lambda_8$ to combine with $\lambda_4^{\prime \prime}$ and $\lambda_5^{\prime \prime}$.
Then, we have focussed on the conservation relationships of 
\begin{eqnarray}
\lambda_1^2 + \lambda_2^2 +\lambda_3^2
&=&
    \cos^2 \left( \frac{\theta_y}{2} \right)    \\
(\lambda_4^{\prime \prime})^2
+
(\lambda_5^{\prime \prime})^2
+
(\lambda_8)^2
&=&
\frac{4}{3}
-
\cos^2 \left( \frac{\theta_y}{2} \right) ,   
\end{eqnarray}
and consider a following non-surjective mapping from SO(6) to SO(5), as we renormalise
\begin{eqnarray}
\overrightarrow{S}
&=&
\left (
  \begin{array}{c}
S_1\\
S_2\\
S_3\\
S_4\\
S_5
  \end{array}
\right)
=
\left (
  \begin{array}{c}
\lambda_1\\
\lambda_2\\
\lambda_3\\
\lambda_4^{\prime \prime \prime}\\
\lambda_5^{\prime \prime \prime}
  \end{array}
\right)
\\
&=&
\left (
  \begin{array}{c}
    \sin \left( \theta_l \right)    
    \cos \left( \phi_l \right)    
    \cos^2 \left( \frac{\theta_y}{2} \right)    \\
    \sin \left( \theta_l \right)    
    \sin \left( \phi_l \right)    
    \cos^2 \left( \frac{\theta_y}{2} \right)    \\
    \cos \left( \theta_l \right)    
    \cos^2 \left( \frac{\theta_y}{2} \right)    \\
\sqrt{\frac{4}{3}-\cos^2 \left( \frac{\theta_y}{2} \right)} 
\cos \left( \phi_y  \right)    
\sin  \left( \theta_y \right)    \\
\sqrt{\frac{4}{3}-\cos^2 \left( \frac{\theta_y}{2} \right)} 
\sin \left( \phi_y  \right)    
\sin  \left( \theta_y \right)       
  \end{array}
\right)
,
\end{eqnarray}
which preserve
\begin{eqnarray}
(\lambda_4^{\prime \prime \prime})^2
+
(\lambda_5^{\prime \prime \prime})^2
&=&
\frac{4}{3}
-
\cos^2 \left( \frac{\theta_y}{2} \right) 
.   
\end{eqnarray}
We also confirm that the renormalised Gell-Mann parameters conserve the norm
\begin{eqnarray}
\sum_{i=1}^5
\left( 
S_i
\right)^2
=
\frac{4}{3}
,
\end{eqnarray}
which is consistent with the constant Casimir operator.
Consequently, expectation values are embedded on a compact Gell-Mann hypersphere of ${\mathbb S}^4$ in SO(5).

\section{Discussions}

\subsection{SU(2)$\times$SU(3) and higher dimensional systems}
So far, we have assumed the ray is polarised such that the polarisation state is fixed.
We can control the polarisation state by a phase-shifter and a rotator.
We have recently proposed a Poincar\'e rotator, which allows an arbitrary rotation of polarisation state by realising SU(2) rotations in a combination of half- and quarter-wave plates and phase-shifters \cite{Saito21f,Saito22g,Saito22h,Saito22i}.
If we use the Poincar\'e rotator for the ray with SU(3) states of vortices under certain polarisation, we can realise the SU(2)$\times$SU(3), since spin and orbital angular momentum are different quantum observables, such that a general state is made of a direct product state for spin and orbital angular momentum.
We can also envisage to realise a state made by a sum of these states with different spin and orbital angular momentum.
For example, if we realise the SU(2) state of left- and right-vortices and assign horizontally and vertically polarised states, respectively, we can realise both singlet and triplet states by controlling the phase among 2 different many-body states.

\subsection{Cavity QCD and photonic mesons}
It is well established a photonic crystal is an excellent test bed to explore a cavity Quantum Electro-Dynamics (QED) in an artificial environment \cite{Joannopoulos08}.
Here we consider an analogue to a cavity QED as a cavity QCD.
We can construct a one-dimensional cavity, for example, as a Fabry-Perot interferometer, where $|{\rm L}\rangle$, $|{\rm R}\rangle$, and $|{\rm O}\rangle$ states are realised.
The ray is propagating in the cavity along $z$, and reflected back to propagate along the opposite direction of $-z$.
The chiralities of spin and orbital angular momentum will be reversed upon reflections \cite{Max99,Jackson99,Yariv97,Gil16,Goldstein11,
Hecht17,Pedrotti07,Saito20a,Saito20b,Saito20c,Saito20d}, such that the state along $-z$ would be a conjugate state to the state along $z$.
Consequently, we will be able to construct multiplets similar to mesons, made of quarks and anti-quarks  \cite{Pfeifer03,Sakurai14,Georgi99,Weinberg05}.
For quarks, an individual quark is very difficult to be observed in experiments due to the strong confinements in composite materials of mesons and baryons.
On the other hand, we expect an opposite behaviour, since photons trapped inside the cavity are difficult to observe as is, while photons escaping from the cavity are observed and analysed by detectors.
This corresponds to see an individual quark, which is a ray of photons propagating at either $z$ or $-z$.
It is quite hard to observe the composite meson analogue, which is realised inside the cavity and it would be difficult to allocate detectors to see photons propagating along the opposite directions at the same time, which would require a transparent detector.
But, it would not be essential to observe within the cavity, since we can examine the state inside the cavity from the photons escaping from the both ends.
The cavity QCD experiments will open to explore SU(3) and SU(2)$\times$SU(3) multiplets in a standard photonic experimental set-up.
If we distinguish each polarisation state with different orbital angular momentum as an individual orthogonal state, we can also explore SU(6) states, for example, and it will also be possible to investigate how symmetry breaking from SU(6) to SU(2)$\times$SU(3) affect the photonic states by observing corresponding expectation values of generators of rotations in a higher dimensional space.
Another remarkable difference of proposed photonic systems with quarks is quantum statistics; quarks are fermions and photons are bosons.
Our analysis is quite primitive, such that some of our ideas could be applicable to femionic systems.
However, coherent photons out of lasers are quite easy to treat due to the technological advances, while macroscopic number of photons are coherently degenerate, which would be ideal for experiments, which require coherent interference.
As we have seen above, phases and amplitudes of a wavefunction determine the crucial Gell-Mann parameters, similar to Stokes parameters for polarisation.
Polarisation is a macroscopic manifestation of the nature of spin for photons, represented on the Poincar\'e sphere.
A similar argument will hold for orbital angular momentum of coherent photons and Gell-Mann hypersphere can play a similar role to clarify the SU(3) states for photons.

\subsection{Correlation between SU(n) and SO($n^2-1$)}

Finally, we would like to discuss the relationship between SU(n) wavefunctions and expectation values in SO($n^2-1$).
It is well known that the SU(2) wavefunction for spin is related to spin average values in SO(3), and therefore, the rotation in SU(2) is linked to the corresponding rotation in SO(3).
This fact is also explained by the relationship SU(2)/${\mathbb Z}_2 \cong$SO(3), claiming that the SU(2) is the twofold coverage of SO(3).
In this paper, we have discussed the relationship betweeen SU(3) and SO(8).
More generally, a quantum mechanical average of an generator in SU(n) is related to a rotation in SO($n^2-1$) using adjoint representation of the $\mathfrak{su}(n)$ algebra.

We assume a generator of rotation in SU(n) is $\hat{X}_a$ and the commutation relationship is $[\hat{X}_a, \hat{X}_b]=i\sum_c f_{abc}\hat{X}_c$ \cite{Georgi99}.
In the above example of SU(3), this corresponds to $\hat{X}_a=\hat{\lambda}_a$. 
We consider an initial SU(n) state of $|{\rm I} \rangle$ will be rotated by an exponential map of $\exp(-i \hat{X}_a \theta)$ with the angle of $\theta$ to be the final state
\begin{eqnarray}
|{\rm F} \rangle
&=&
{\rm e}^{-i \hat{X}_a \theta}
|{\rm I} \rangle.
\end{eqnarray}
Then, we consider how an average expectation value of $\hat{X}_b$ in the initial state $\langle 
\hat{X}_b \rangle_{\rm I}= \langle {\rm I} | \hat{X}_b |{\rm I} \rangle$ is transferred in the final state as 
\begin{eqnarray}
\langle 
\hat{X}_b
\rangle_{\rm F}
&=&
\langle {\rm F} |
\hat{X}_b
|{\rm F} \rangle.
\nonumber
\\
&=&
\langle {\rm I} |
{\rm e}^{i \hat{X}_a \theta}
\hat{X}_b
{\rm e}^{-i \hat{X}_a \theta}
|{\rm I} \rangle.
\nonumber
\\
&\approx&
\langle {\rm I} |
(1+i \hat{X}_a \theta)
\hat{X}_b
(1-i \hat{X}_a \theta)
|{\rm I} \rangle
+{\mathcal O}(\theta^2)
\nonumber
\\
&\approx&
\langle {\rm I} |
(\hat{X}_b+i \theta [\hat{X}_a,\hat{X}_b])
|{\rm I} \rangle
+{\mathcal O}(\theta^2)
\nonumber
\\
&\approx&
(\delta_{bc}-\sum_c f_{abc} \theta)
\langle \hat{X}_c \rangle_{\rm I}
+{\mathcal O}(\theta^2)
\nonumber
\\
&\approx&
\sum_c
\left( 
{\rm e}^{- \hat{F}_{a} \theta}
\right)_{bc}
\langle \hat{X}_c \rangle_{\rm I}
+{\mathcal O}(\theta^2)
,
\end{eqnarray}
where we have assumed $\theta$ is infinitesimally small and considered only the first order in the expansion, and $\hat{F}_{a}$ is an adjoint operator, whose matrix element becomes $(\hat{F}_{a})_{bc}=f_{abc}$, which is a matrix of $(n^2-1)\times(n^2-1)$.
Therefore, the rotation of the wavefunction in SU(n) becomes the rotation of the corresponding expectation value in SO($n^2-1$), as we expected.

We have also checked its validity in the second order of $\theta$ as
\begin{eqnarray}
{\mathcal O}(\theta^2)
&=&
-\frac{\theta^2}{2}
\langle {\rm I} |\hat{X}_a^2 \hat{X}_b | {\rm I} \rangle
+
\theta^2
\langle {\rm I} |\hat{X}_a \hat{X}_b \hat{X}_a | {\rm I} \rangle
\nonumber
\\
&&
-\frac{\theta^2}{2}
\langle {\rm I} |\hat{X}_b^2 \hat{X}_a | {\rm I} \rangle
\nonumber
\\
&=&
-\frac{\theta^2}{2}
\langle {\rm I} |
(\hat{X}_a^2 \hat{X}_b
-2 \hat{X}_a \hat{X}_b \hat{X}_a
+\hat{X}_b^2 \hat{X}_a
 | {\rm I} \rangle
 \nonumber \\
&=&
-\frac{\theta^2}{2}
\langle {\rm I} |
(
\hat{X}_a^2 \hat{X}_b
\nonumber \\ 
&&
-2 \hat{X}_a(\hat{X}_a\hat{X}_b -i\sum_cf_{abc}\hat{X}_c)
\nonumber \\ 
&&
+\hat{X}_a^2 \hat{X}_b
-i\sum_cf_{abc}(\hat{X}_a\hat{X}_c+\hat{X}_c\hat{X}_a)
) | {\rm I} \rangle
\nonumber \\ 
&=&
-\frac{\theta^2}{2}
\langle {\rm I} |
(
2i\sum_cf_{abc}\hat{X}_a\hat{X}_c)
) | {\rm I} \rangle
\nonumber
\\
&=&
-i\frac{\theta^2}{2}
\sum_c
\langle {\rm I} |
(
f_{abc}\hat{X}_a\hat{X}_c
+
f_{cba}\hat{X}_c\hat{X}_a
)
) | {\rm I} \rangle
\nonumber
\\
&=&
-i\frac{\theta^2}{2}
\sum_c
f_{abc}
\langle {\rm I} |
[
\hat{X}_a,\hat{X}_c
] | {\rm I} \rangle
\nonumber
\\
&=&
\frac{\theta^2}{2}
\sum_{cd}
f_{abc}
f_{acd}
\langle {\rm I} |
\hat{X}_d
| {\rm I} \rangle
\nonumber
\\
&=&
\frac{\theta^2}{2}
\sum_{c}
(\hat{F}_a^2)_{bc}
\langle {\rm I} |
\hat{X}_c
| {\rm I} \rangle
,
\end{eqnarray}
and therefore, the above formula is also valid in the second order.
Actually, this is the reflection of the differentiability of the Lie group, which was originally called as an infinitesimal group.
Once a formula is derived in the infinitesimal small value, it is straightforward to extend it to the finite value.
In our case, we can repeat the infinitesimal amount of rotation with the angle of $\theta/N$, while we can repeat N times, and we take the limit $N \rightarrow \infty$ as
\begin{eqnarray}
\langle 
\hat{X}_b
\rangle_{\rm F}
&=&
\lim_{N\rightarrow \infty}
\sum_c
\left( 1- \hat{F}_{a} \frac{\theta}{N} \right)_{bc}^{N}
\langle \hat{X}_c \rangle_{\rm I}
\\
&=&
\sum_c
\left( 
{\rm e}^{- \hat{F}_{a} \theta}
\right)_{bc}
\langle \hat{X}_c \rangle_{\rm I}
.
\end{eqnarray}
Therefore, we have proved that the quantum mechanical rotation of the wavefunction in SU(N), which is given by ${\mathbb C}^n$ on ${\mathbb S}^(n-1)$ upon the normalisation, will rotate the expectation value of the generator, which is a vector of ${\mathbb R}^{n^2-1}$ , in SO($n^2-1$), using the adjoint operator of $\mathfrak{su}(3)$ Lie algebra.

\section{Conclusion}
We have proposed to use photonic orbital angular momentum for exploring the SU(3) states as a photonic analogue of QCD.
We have shown that the 8-dimensional Gell-Mann hypersphere in SO(8) characterises the SU(3) state, made of left- and right-vortexed photons and no-vortexed photons.
There are several ways to visualise the Gell-Mann hypersphere, and we have calculated expectation values for orbital angular momentum and defined hyperspin to represent the coupling between vortexed and no-vortexed states, which could be shown on 2 Poincar\'e sphere or 1 hypersphere in SO(6) or SO(5).
We believe the proposed superposition state of photons are useful to explore photonic many-body states to have some insights on the nature of the symmetries in photonic states.

\section*{Acknowledgements}
This work is supported by JSPS KAKENHI Grant Number JP 18K19958.
The author would like to express sincere thanks to Prof I. Tomita for continuous discussions and encouragements. 


\bibliography{SU3}

\end{document}